\documentclass[12pt]{article}
\usepackage{amsmath,amsfonts,amssymb}
\usepackage{graphicx}

\newcommand{\be}{\begin{equation}}
\newcommand{\ee}{\end{equation}}
\newcommand{\ba}{\begin{eqnarray}}
\newcommand{\ea}{\end{eqnarray}}

\begin{document}
\newcommand{\todo}[1]{{\em \small {#1}}\marginpar{$\Longleftarrow$}}   
\newcommand{\labell}[1]{\label{#1}\qquad_{#1}} 

\rightline{DCPT-11/35}   
\vskip 1cm 

\begin{center} {\Large \bf Holography for asymptotically locally Lifshitz spacetimes}
\end{center} 
\vskip 1cm

\renewcommand{\thefootnote}{\fnsymbol{footnote}} 
\centerline{Simon F. Ross\footnote{s.f.ross@durham.ac.uk} }
\vskip .5cm 
\centerline{\it Centre for Particle Theory, Department of Mathematical Sciences}
\centerline{\it Durham University, South Road, Durham DH1 3LE, U.K.}

\setcounter{footnote}{0}   
\renewcommand{\thefootnote}{\arabic{footnote}}

\begin{abstract}
We give a definition of asymptotically locally Lifshitz spacetimes, with boundary data appropriate for a non-relativistic theory on the boundary. Solutions satisfying these boundary conditions are constructed in an asymptotic expansion. We identify the boundary data with sources for dual field theory operators, and give a prescription for calculating the one-point functions of the field theory operators (including the stress tensor) in the presence of arbitrary sources. The divergences in these one-point functions can be cancelled by holographic renormalization, adding counterterms which are local functions of the boundary data. 
\end{abstract}

\section{Introduction}

The use of gravitational duals to study strongly-coupled field theories \cite{Maldacena:1997re,Aharony:1999ti} has provided a unique tool which has shed light on a number of important questions, particularly concerning thermodynamics and theories at finite density, which are related to black holes in the bulk. This work has been extended  over the last few years to include applications to field theories of interest to condensed matter physics (see \cite{Hartnoll:2009sz,McGreevy:2009xe} for useful reviews). The application to condensed matter throws up new questions; to address these and fully realise the potential of the holographic approach, we need to extend our understanding of the bulk gravitational theories and the holographic dictionary. An important example of this is the appearance of anisotropic scaling symmetries in condensed matter. The low-energy physics near a phase transition may be invariant under $t \to \lambda^z t, x^i \to \lambda x^i$, where $z$ is referred to as the dynamical exponent.  We follow  \cite{Kachru:2008yh} in referring to a field theory which has this scaling symmetry (and no boost symmetry) as a Lifshitz field theory. Understanding such symmetries holographically requires us to go beyond the familiar context of asymptotically anti-de Sitter spacetimes on the gravitational side.

A holographic duality for these field theories was proposed in \cite{Kachru:2008yh}. The proposal is that the dual of the field theory vacuum has a bulk metric
\begin{equation} \label{lif} 
ds^2 = -r^{2z} dt^2 + r^{2} d\vec{x}^2 + L^2 \frac{dr^2}{r^2},
\end{equation}
where $L^2$ represents the overall curvature scale, and the spacetime has $d+1$ dimensions, so there are $d_s = d-1$ spatial dimensions $\vec{x}$.\footnote{We will mostly focus on the case $d_s = 2$.} This metric is referred to as a Lifshitz geometry. Such a metric can be realised as a solution in a variety of bulk gravitational theories with different matter content. In \cite{Kachru:2008yh}, the bulk theory involved two $p$-form fields with a Chern-Simons coupling. A simpler theory with a massive vector (which is on-shell equivalent to the previous theory) was introduced in \cite{Taylor:2008tg}. More recently, \eqref{lif} was realised as a solution in string theory in a number of different truncations \cite{Balasubramanian:2010uk,Gregory:2010gx,Donos:2010tu,Donos:2010ax,Cassani:2011sv} (see \cite{Hartnoll:2009ns} for earlier attempts). In this paper, we will focus on the massive vector theory of \cite{Taylor:2008tg}, which provides the simplest context for studying this geometry. The focus is on the differences in structure between geometries of the form \eqref{lif} and the AdS case, so the key results will not depend too heavily on the specific form of the matter considered. 

In the AdS context, there is a well-developed holographic dictionary, starting from the work of \cite{Gubser:1998bc,Witten:1998qj}. This relates asymptotic boundary conditions for bulk fields to the sources for dual operators in the field theory description. The asymptotic boundary data for the metric is specified by the induced metric on the conformal boundary of the spacetime. The results of \cite{FG,Graham:1999jg} show crucially that at least in a neighbourhood of infinity, there is a bulk solution for any arbitrary boundary metric, so we can consider the correspondence for the field theory on an arbitrary background. 

The correlation functions for operators in the CFT are then obtained in the saddle-point approximation by considering the variational derivative of the bulk action with respect to the sources. When this is applied to the naive bulk action, these correlation functions contain divergences. These can be removed by the process of holographic renormalization, adding appropriate local boundary counterterms to the bulk action to ensure that it defines a good variational principle for the asymptotic boundary conditions \cite{Balasubramanian:1999re,Henningson:1998gx}. An elegant approach based on Hamiltonian evolution in the radial direction and a functional differential was developed by \cite{Papadimitriou:2004ap,Papadimitriou:2004rz}. 

The aim of the present paper is to develop a general holographic dictionary for the Lifshitz case. Starting with \cite{Kachru:2008yh}, there has been extensive work on the calculation of correlation functions from perturbations in the bulk; but so far, this work has only considered spacetimes which asymptotically approach \eqref{lif}, treating changes in the asymptotic boundary data only perturbatively. Here we extend this to consider arbitrary boundary data, constructing metrics which approach \eqref{lif} only locally. In the AdS case, this generalisation is achieved by considering an arbitrary conformal class of metrics on the conformal boundary of the spacetime. In addition to enabling us to treat field theories in curved backgrounds, this also provides a more comprehensive (and less coordinate-dependent) perspective on the holographic dictionary, allowing us to understand its essential features. 

In the Lifshitz case, the spacetime \eqref{lif} does not have a conformal boundary, as the timelike direction is treated differently from the spatial ones. Thus, while a similar correspondence should exist, we need to develop a new approach to describe it.  Building on the work of \cite{Ross:2009ar} (and similar observations in \cite{Horava:2009vy}), in section \ref{all}  a definition of asymptotically locally Lifshitz spacetimes is given, in terms of appropriate boundary conditions on a set of frame fields describing the bulk geometry. This enables a coordinate-invariant treatment of the different scaling of the timelike direction. The rescaled boundary values of the frame fields define the boundary data for the geometry; they can be thought of as defining a Galilean geometrical structure on the spacetime boundary. If we impose a further restriction, the boundary admits a notion of absolute time.

In section \ref{cft}, the interpretation of the boundary data from the dual field theory point of view is discussed. Some general expectations for the behaviour of the stress tensor complex in a Lifshitz theory are reviewed. Then following \cite{Ross:2009ar}, a prescription for calculating this stress tensor complex is given, considering the variation of an appropriate bulk action with respect to variations in the boundary data for the frames, holding the boundary value of the matter field(s) with tangent space indices fixed. 

In section \ref{mvm}, we specialise the discussion to the massive vector model of \cite{Taylor:2008tg}, to enable discussion of more dynamical issues. We review esults of \cite{Ross:2009ar} on the solution of the linearized equations of motion on the background \eqref{lif}. We discuss some of the open problems arising from that work, some of which will be addressed here and some of which remain for future work. In particular, we find that a modification of the boundary conditions is required for $z > 4$; we treat this modification in the linearized regime, but leave a full exploration of the modified boundary conditions for future work.

In section \ref{eom}, the solution of the equations of motion in a large-distance expansion is considered. We find that asymptotically locally Lifshitz solutions exist for arbitrary boundary data for $z < 2$. For $z \geq 2$, there are restrictions on the boundary data, setting to zero the sources for the irrelevant operators. There are then asymptotically locally Lifshitz solutions for arbitrary sources for the relevant and marginal operators. These are key results, providing an analogue of the Fefferman-Graham expansion \cite{FG} for asymptotically Lifshitz boundary conditions.

Finally, the problem of holographic renormalization is considered in section \ref{ct}. We obtain counterterms rendering the expectation value of the stress tensor finite for arbitrary asymptotically locally Lifshitz spacetimes. The counterterms are obtained using an analogue of the approach introduced for AdS in \cite{Papadimitriou:2004ap,Papadimitriou:2004rz}, organising the calculation in eigenvalues of a dilatation operator. This approach efficiently evaluates the counterterms; this is illustrated by an explicit calculation of the first few counterterms. For $z >2$, it was observed in \cite{Ross:2009ar} that there were divergences in components of the stress tensor even when the sources were set equal to zero; some of the response functions cause divergences in other expectation values. Here we will see that the form of the counterterms is uniquely fixed by requiring cancellation of the divergence coming from the sources, but that this same term also cancels the divergence from the response functions, so that these are ``pseudo-non-local'' in the terminology of \cite{vanRees:2011fr}. 

\textit{Note added:} After the appearance of this paper, partial results on holographic renormalisation for asymptotically Lifshitz spacetimes were also reported in \cite{BBH,MM}.

\section{Asymptotically locally Lifshitz spacetimes}
\label{all}

We want to consider spacetimes which asymptotically approach a metric which can locally be written in the form \eqref{lif}. This condition can be most easily implemented by writing the geometry in terms of a set of frame fields. We write the spacetime metric as 
\begin{equation}
ds^2 = \eta_{MN} e^{(M)} e^{(N)} = \eta_{AB} e^{(A)} e^{(B)} + (e^{(r)})^2, 
\end{equation}
where spacetime frame indices are written $M, N = 0,1, \ldots d$, while frame indices omitting the radial direction are $A, B = 0, \ldots, d_s$. We will also sometimes use spatial frame indices $I, J$ running over $1, \ldots, d_s$ (recall $d_s = d-1$ is the number of spatial dimensions). Similarly spacetime coordinate indices are $\mu, \nu = t, x_i,  r$ and boundary coordinate indices $\alpha, \beta = t, x_i$. We partially fix the gauge freedom by choosing a Gaussian normal radial coordinate such that $e^{(r)} = L r^{-1} dr$, and $e^{(A)}$ have no radial components. 

For the metric \eqref{lif}, there is a natural set of frame fields where $e^{(0)} = r^{z} dt$, $e^{(I)} = r dx^i$.  This motivates defining asymptotically locally Lifshitz boundary conditions by requiring that the spacetime admit a choice of frames $e^{(A)}$ such that as $r \to \infty$,
\begin{equation} \label{frame}
\boxed{e^{(0)}_\alpha = r^{z} \hat e^{(0)}_\alpha(r, x^\alpha), \quad e^{(I)}_\alpha = r \hat e^{(I)}_\alpha(r, x^\alpha),}
\end{equation}
where $\hat e^{(0)}_\alpha(r, x^\alpha)$, $\hat e^{(I)}_\alpha(r, x^\alpha)$ have some finite non-degenerate limits as $r \to \infty$, which we will also sometimes call $\hat e^{(0)}_\alpha(x^\alpha)$, $\hat e^{(I)}_\alpha(x^\alpha)$.  That is, we simply replace the coordinate frame fields $dt$, $dx^i$ in \eqref{lif} by some arbitrary non-degenerate frame basis $\hat e^{(A)}_\alpha(x^\alpha)$. This is the most general boundary condition which allows the metric to locally be written as \eqref{lif} in the asymptotic region.

Note that we have not imposed any specific falloff condition on the subleading parts of $\hat e^{(0)}_\alpha(r, x^\alpha)$, $\hat e^{(I)}_\alpha(r, x^\alpha)$, beyond requiring that they vanish as $r \to \infty$. This is because this boundary condition is meant to be purely kinematical, applying to off-shell fluctuations considered in the discussion of the action as well as to asymptotically locally Lifshitz solutions of any bulk theory. As in the AdS case, in solving the equations of motion in section \ref{eom}, we will find that the dynamics will dictate specific powers of $r$ which appear in the subleading parts of  $\hat e^{(0)}_\alpha(r, x^\alpha)$, $\hat e^{(I)}_\alpha(r, x^\alpha)$ and in the matter fields. But the specific fall-offs will depend on the details of the bulk theory (again, as in the AdS case), so we do not wish to introduce them in our definition of the boundary conditions.

The boundary frame fields $\hat e^{(0)}_\alpha(x^\alpha)$, $\hat e^{(I)}_\alpha(x^\alpha)$ define our boundary data; we will see in the next section how these provide sources for the the stress tensor complex. Geometrically, they define the general background structure appropriate for a non-relativistic field theory on the asymptotic boundary of the spacetime. Such data has been previously defined in terms of what is referred to as a Galilean metric \cite{MTW,Ruede:1996sy}, consisting of a timelike covector (one-form) field, and a degenerate contravariant metric orthogonal to the covector. This is equivalent to our frame data, as the rescaled timelike frame field $\hat e^{(0)}_\alpha(r, x^\alpha)$ provides a distinguished timelike covector on the boundary, while the rescaled inverse metric 
\begin{equation}
\hat g^{\alpha\beta} = \lim_{r \to \infty} r^{2} g^{\alpha\beta} =  \hat{e}^{(I)\alpha} \hat{e}^\beta_{(I)}\end{equation}
provides a degenerate boundary contravariant metric, which is obviously orthogonal to the timelike covector, $\hat g^{\alpha\beta} \hat e^{(0)}_\alpha = 0$. We can also note that the anisotropic scaling implies that the bulk freedom to make local Lorentz transformations of the $e^{(A)}$ is reduced to the freedom to make spatial rotations of the $\hat{e}^{(I)}$, again as we would expect for a non-relativistic theory.

However, in a non-relativistic context it seems natural to require that the boundary have a notion of absolute time. That is, the boundary should have a foliation by a family of surfaces defining ``moments in time''. This is not satisfied for general boundary data satisfying \eqref{frame}: arbitrary $\hat e^{(0)}_\alpha$ identifies only a preferred family of curves parallel to the distinguished vector field $\hat e_{(0)}^\alpha$. To be able to use $\hat e^{(0)}_\alpha$ to identify a preferred foliation by surfaces, we need to require that it is irrotational,
\begin{equation} \label{irr}
\hat{e}^{(0)} \wedge d \hat{e}^{(0)} = 0.
\end{equation}
This condition can be conveniently expressed in terms of the bulk Ricci rotation coefficients $\Omega_{BC}^{\ \ \ A}$, defined by $d e^{(A)} = \Omega_{BC}^{\ \ \ A} e^{(B)} \wedge e^{(C)}$. Requiring  $\hat{e}^{(0)}$ to be irrotational requires the leading term in $\Omega_{IJ}^{\ \ \ 0}$ to vanish asymptotically. If we impose this condition, we can choose coordinates such that $\hat{e}^{(0)} = \chi(t, x^i) dt$ for some function $\chi(t,x^i)$ (at least in an open neighbourhood). The irrotational condition is also a necessary and sufficient condition for the spatial frame fields $\hat{e}^{(I)}$ to be surface forming. Thus, the $\hat{e}^{(I)}$ will describe a non-degenerate $t$-dependent curved metric on the surfaces of constant $t$. 

One might then take \eqref{frame} and \eqref{irr} to define asymptotically locally Lifshitz boundary conditions. However, we will see in the next section that assuming that $\hat e^{(0)}_\alpha$ is irrotational corresponds to setting the sources for the boundary energy flux $\mathcal E^i$ to zero. Thus, to be able to calculate correlation functions involving the energy flux, we need to consider violations of \eqref{irr} at least perturbatively. In the remainder of the paper, we do not in general impose \eqref{irr} (for $z \geq 2$ however, satisfying \eqref{frame} turns out to require that we set the sources for $\mathcal E^i$ to zero, which implies we also satisfy \eqref{irr}).

\section{Field theory sources}
\label{cft}

We now discuss the interpretation of this boundary data in terms of sources for the field theory stress tensor. We make the central assumption that as in AdS/CFT, asymptotic boundary data for bulk fields should be interpreted as sources for dual operators in the dual field theory. In the case of the metric (frame fields), the appropriate dual operator should be the stress tensor complex, so we first review a few elements of the field theory expectations for its components. 

In a non-relativistic theory, the tensor complex should consist of an energy density $\mathcal E$ and an energy flux $\mathcal E^i$, satisfying a conservation equation (in flat space)
\begin{equation} \label{cons1}
\partial_t \mathcal E + \partial_i \mathcal E^i = 0,
\end{equation}
and a momentum density $\mathcal P_i$ and a spatial stress tensor $\Pi_i^j$, satisfying the conservation equations (in flat space)
\begin{equation} \label{cons2}
\partial_t \mathcal P_i + \partial_j  \Pi_i^{\ j} = 0. 
\end{equation}
As in the relativistic case, invariance under the anisotropic scaling symmetry $t \to \lambda^z t, x^i \to \lambda x^i$ implies a  tracelessness condition $z \mathcal E + \Pi_i^{\ i} = 0$.  The key difference from the relativistic case is that the momentum density and energy flux in a non-relativistic theory are independent quantities. 

In a Lifshitz theory, there are also some differences in the operator dimensions for these quantities. These do not seem to have been discussed earlier in the literature, so it is worth considering them here. Since $\mathcal E$ is the energy density, and the energy is dimension $z$ with respect to the anisotropic scaling symmetry, it will have dimension $z+d_s$. This is the marginal dimension for a Lifshitz theory, as it matches the scaling of the volume element. Thus, energy density is a marginal operator, as in the relativistic case. The conservation equation then implies that the energy flux has dimension $2z+d_s-1$. This implies that it is an irrelevant operator for $z>1$. Thus, deforming the theory by adding sources for the energy flux will change the UV theory, driving it away from the fixed point, quite unlike the relativistic case.

The momentum density $\mathcal P_i$ will have dimension $d_s+1$, as momentum is dimension $1$ with respect to the anisotropic scaling symmetry. This is relevant for $z>1$. Since $\mathcal P_i$ has dimension $d_s+1$, the conservation equation implies that $\Pi_{\ i}^j$ has dimension $z+d_s$, indicating that it is also marginal. This is consistent with the condition $z \mathcal E + \Pi_{\ i}^i = 0$, which requires that $\Pi_{\ i}^j$ must have the same dimension as $\mathcal E$. 

The fact that $\mathcal P_i$ is relevant has several interesting consequences. First, it implies that we can generate non-trivial renormalization group flows to the IR by adding sources for the momentum density. Secondly, the source for this operator will have dimension $z-1$, so for $z > d_s+2$, the dimension of the source is greater than the dimension of the operator. From the holographic point of view, this suggests that the boundary data associated with this source will fall off more quickly at large $r$ than the mode giving the source. We will see this explicitly in the linearised analysis in the next section. This is reminiscent of the behaviour of scalar fields in the alternative quantization of \cite{Klebanov:1999tb}. A similar $z$-dependent crossover between the source and the expectation value was seen for a Maxwell field on a Lifshitz background in \cite{Hartnoll:2009ns}. The scalar operator $\mathcal P_i \mathcal P^i$ becomes relevant for $z > d_s+2$ with the fixed source boundary conditions, implying that there is then a relevant deformation of the theory which preserves the spatial rotation symmetry. The natural endpoint for this flow is a Lifshitz theory with fixed $\mathcal P_i$ boundary conditions.

How are the components of the stress tensor complex calculated holographically? As argued in \cite{Ross:2009ar}, the appropriate dictionary is to identify the timelike frame field $\hat e^{(0)}$ as the boundary data supplying the sources for $\mathcal E$, $\mathcal E^i$, and the other frame fields $\hat e^{(I)}$ as the boundary data supplying the sources for $\mathcal P_i$, $\Pi^i_j$. More explicitly, assume we have a bulk action $S[\hat e^{(A)}, \psi]$ (where $\psi$ denotes whatever matter fields we consider), which is finite on-shell and provides a good variational principle for our boundary conditions, so $\delta S = 0$ for variations satisfying $\delta \hat e^{(A)} = 0$.\footnote{This is a non-trivial assumption, but section \ref{ct} shows that we can provide such an action by holographic renormalization.} Then writing
\begin{equation} \label{fvar}
\delta S = \int \hat \epsilon (T^\alpha_{\ B}  \delta \hat e^{(B)}_\alpha  + O_\psi \delta \hat \psi),
\end{equation}
where $\hat\epsilon$ is the coordinate-invariant volume density given by $\hat \epsilon = \hat{e}^{(0)} \wedge \hat{e}^{(1)} \wedge ... \wedge \hat{e}^{(d_s)}$, we identify $T^\alpha_{\ 0}$ with the vacuum expectation value (vev) of the energy density $\mathcal E$ and the energy flux $\mathcal E^i$, and $T^\alpha_{\ I}$ with the vev of the  momentum density $\mathcal P_i$ and the stress tensor $\Pi^i_{\ j}$. In accordance with the usual holographic dictionary, we also identify $O_\psi$ with the vevs of operators dual to the matter fields. Note that to have the Lifshitz solution \eqref{lif}, the bulk theory will need to include vector or tensor fields which are non-zero in the Lifshitz geometry, which dynamically select the timelike direction as different from the spacelike ones. Following \cite{Hollands:2005ya,Ross:2009ar} the $\psi$ are then understood to include the tangent space components of these vector or tensor fields. That is, the variation of the frame fields in \eqref{fvar} is with the tangent space components of these quantities fixed, rather than the spacetime components. This is an essential ingredient to ensure that $T^i_{\ 0}$ and $T^t_{\ I}$ are different, so that they can be identified with the distinct components of the stress tensor complex \cite{Guica:2010sw}. It can also be understood physically by arguing that we take the variation with the tangent space indices fixed so that the alignment between the distinguished frame field $e^{(0)}$ and the matter fields that are responsible for singling it out is maintained as we do this variation. 

One simple check of this prescription is that these quantities are conserved as required from the field theory point of view. If we assume that the action $S$ is invariant under diffeomorphisms of the boundary coordinates, then as shown in the asymptotically AdS context in \cite{Hollands:2005ya}, this diffeomorphism symmetry implies a conservation equation, 
\begin{equation} \label{cons}
\nabla_\alpha T^\alpha_{\ \beta} - O_\psi \nabla_\beta \hat \psi = 0,
\end{equation}
so $T^\alpha_{\ \beta}$ is conserved up to the presence of sources. This is the analogue of \eqref{cons1}, \eqref{cons2} for general boundary sources. 

Note that the sources for the energy flux $\mathcal E^i$ are $\hat e^{(0)}_i$, the $dx^i$ components of the timelike frame field. Thus imposing the irrotational condition, which allows us to set $\hat e^{(0)}_i = 0$ by choice of coordinate system, is setting these sources to zero (up to diffeomorphism). Since $\mathcal E^i$ is an irrelevant operator in the field theory, it is not surprising that allowing arbitrary sources for it leads to a qualitative change in the UV behaviour, violating what we may want to call asymptotically Lifshitz boundary conditions. The precise form of this change is however a little unusual. At least for small $z$, it is not that some field is growing too quickly (as we will see again in the next section), but rather there is an obstruction to defining an absolute time on the boundary.

\section{Massive vector theory}
\label{mvm}

Our discussion so far has been very general; we will now turn to the specific example of the massive vector theory of \cite{Taylor:2008tg}, in four bulk spacetime dimensions (so $d_s= 2$), so that we can consider more dynamical issues. In this section, we will give the definition of this theory and review the previous work of \cite{Ross:2009ar} on the linearized equations of motion on the background \eqref{lif}. We will also comment on issues to do with the boundary conditions in light of this linearised analysis. In the next section we construct solutions for arbitrary boundary data in an asymptotic analysis. While we focus on the massive vector model as a specific example, the extension of this analysis to for example the string theory truncations of \cite{Balasubramanian:2010uk,Donos:2010tu,Donos:2010ax,Gregory:2010gx} is in principle straightforward.  

The bulk spacetime action for the massive vector theory is 
\begin{equation} \label{Dact}
S =-\frac{1}{16 \pi G_4} \int d^4x \sqrt{-g} (R - 2\Lambda -
\frac{1}{4} F_{\mu\nu} F^{\mu\nu} - \frac{1}{2} m^2 A_\mu A^\mu) -
\frac{1}{8 \pi G_4} \int d^3 \xi \sqrt{-h} K,
\end{equation}
where we have included the Gibbons-Hawking surface term, so that this is a well-behaved action principle for manifolds with boundary with Dirichlet boundary conditions. The equations of motion for this theory are
\begin{equation} \label{Eeqn}
R_{\mu\nu} = \Lambda g_{\mu\nu} + \frac{1}{2} F_{\mu\lambda} F_\nu^{\ \lambda} -
\frac{1}{8} F_{\lambda\rho} F^{\lambda \rho} g_{\mu\nu}  +
\frac{1}{2} m^2 A_\mu A_\nu
\end{equation}
and
\begin{equation} \label{Meqn}
\nabla_\mu F^{\mu\nu} = m^2 A^\nu.
\end{equation}
To have the solution \eqref{lif}, we choose $\Lambda = -\frac{1}{2L^2}(z^2+z+4)$ and $m^2 L^2=2z$. Then the theory has a solution with metric \eqref{lif} and 
\begin{equation}
A = \alpha r^{z} dt = \sqrt{\frac{2(z-1)}{z}} r^{z} dt.
\end{equation}

\subsection{Linearized analysis}
\label{lin}

The linearized analysis in \cite{Ross:2009ar} found a general solution of the  linearized equations of motion around the pure Lifshitz background \eqref{lif}.\footnote{A similar analysis also appeared in \cite{Bertoldi:2009vn}. Recent extensions appeared in \cite{Zingg:2011cw,BBH}.} From this linearized analysis, one can identify the modes corresponding to an infinitesimal change in the sources given by the boundary values of $\hat{e}^{(A)}$, and the modes corresponding to the part of the solution of the equations of motion which is not locally determined in terms of these boundary values. The essential features can be seen from the analysis of linearized perturbations which are constant along the boundary directions. Written in terms of frame fields, the results of \cite{Ross:2009ar} are that 
\begin{equation}
e^{(0)} = r^{z} ( 1 + \frac{1}{2} f) dt  + r v_{1i} dx^i, \quad e^{(i)} = r^{z} v_2^i dt + r (\delta^i_{\ j} + \frac{1}{2} k^i_{\ j}) dx^j,   
\end{equation}
where 
\begin{equation}
f = c_4 + \frac{4}{z+2} c_1 r^{-(z+2)} + 2
\frac{(5z-2-\beta_z)}{(z+2+\beta_z)}
c_2 r^{-\frac{1}{2}(z+2+\beta_z)}  -2
\frac{(5z-2+\beta_z)}{(z+2-\beta_z)}
c_3 r^{-\frac{1}{2}(z+2-\beta_z)},
\end{equation}
\begin{eqnarray}
v_{1i}(r) &=& c_{1i} r^{(z-1)} + c_{2i} r^{-3} + c_{3i} r^{-(2z+1)},
\\
v_{2i}(r) &=& c_{4i} r^{(1-z)} + \frac{(z^2-4)}{z(z-4)} c_{2i} r^{-3} +
\frac{3z}{(z+2)}  c_{3i} r^{-(2z+1)}, \nonumber
\end{eqnarray}
and
\begin{equation}
k_{ij} = \delta_{ij} k + k_{ij}^{TT}, 
\end{equation}
with 
\begin{equation}
k = c_5 +  \frac{2}{(z+2)} c_1 r^{-(z+2)} - 2
\frac{(3z-4-\beta_z)}{(z+2+\beta_z)}
c_2 r^{-\frac{1}{2}(z+2+\beta_z)} +2
\frac{(3z-4+\beta_z)}{(z+2-\beta_z)}
c_3 r^{-\frac{1}{2}(z+2-\beta_z)},
\end{equation}
and
\begin{equation}
k^{TT}_{xx} = k^{TT}_{yy} = t_{d1} + t_{d2} r^{-(z+2)}, \quad k^{TT}_{xy} = k^{TT}_{yx} = t_{o1} + t_{o2} r^{-(z+2)}. 
\end{equation}
The massive vector field was taken to be of the form $A_A = \delta_A^0 A_0$ by choice of frame, choosing the frame vector $e_{(0)}$ to be aligned with the part of the vector field parallel to the boundary, and $
A_0 = \alpha + j, 
$ with
\begin{equation}
j = -\frac{(z+1)}{(z-1)} c_1 r^{-(z+2)} -
\frac{(z+1)}{(z-1)} c_2 r^{-\frac{1}{2}(z+2+\beta_z)} +
\frac{(z+1)}{(z-1)} c_3 r^{-\frac{1}{2}(z+2-\beta_z)}. 
\end{equation}
Here $\beta_z^2 = 9z^2 - 20z + 20 = (z+2)^2 + 8(z-1)(z-2)$. Note these results are for generic values of $z$. For specific even integer values, there will be logarithmic solutions; see \cite{Ross:2009ar} for details. Also, here we have redefined $v_{1i}$, $v_{2i}$ compared to \cite{Ross:2009ar} to highlight the field theory interpretation of the bulk modes. We adjusted the explicit powers of $r$ in front of $v_{1i}$, $v_{2i}$ in $e^{(A)}$ so that all of $f, j, v_{1i}, v_{2i}, k_{ij}$ are unchanged under the dilatation isometry $r \to \lambda^{-1}r $, $t \to \lambda^z t$, $\vec{x} \to \lambda \vec{x}$ of the background \eqref{lif}. This ensures that the scaling dimension of the different modes in $f, j, v_{1i}, v_{2i}, k_{ij}$ can simply be read off from the powers of $r$ associated with them.

This is a solution of the linearized equations of motion for arbitrary constant values of the coefficients. If we promote the coefficients to functions of the boundary coordinates, these are still solutions to leading order in an expansion in $r$, but there will be further subleading terms determined in terms of the derivatives of the coefficients, as described in detail in \cite{Ross:2009ar}.

The physical significance of these modes can be readily identified:
\begin{itemize}
\item The modes $c_4$, $c_{1i}$ are changes of $\lim_{r \to \infty} \hat{e}^{(0)}$. Accordingly, these are interpreted as sources for $\mathcal E$, $\mathcal E^i$. 
\item The modes $c_{4i}$, $c_5$, $t_{d1}$, $t_{o1}$  are changes of $\lim_{r \to \infty} \hat{e}^{(I)}$. Accordingly, these are interpreted as sources for $\mathcal P_i$, $\Pi^i_{\ j}$. 
\item The modes $c_1$, $t_{d2}$, $t_{o2}$ have scaling dimension $z+2$. They were shown to give finite contributions to the expectation values of $\mathcal E$, $\Pi^i_{\ j}$ in \cite{Ross:2009ar}.
\item The modes $c_{2i}$ have scaling dimension $3$ and were shown to give finite contributions to the expectation values of $\mathcal P_i$ in \cite{Ross:2009ar}.
\item The modes $c_{3i}$ have scaling dimension $2z+1$ and were shown to give finite contributions to the expectation values of expectation values of $\mathcal E^i$ in \cite{Ross:2009ar}.
\item The remaining modes $c_3$, $c_2$ can be interpreted as the source and expectation value of an operator $\mathcal O_\psi$ associated with the massive vector field. From their scaling dimensions, we can read off the dimension of this operator, $\Delta_\psi = \frac{1}{2} (z+2 + \beta_z)$. Note that this operator will be relevant for $z <2$, and irrelevant for $z > 2$. As one might expect for an irrelevant mode, the source $c_3$ makes contributions to $e^{(0)}$, $e^{(I)}$ which violate the boundary conditions when $z>2$. The mode $c_2$ was shown to give finite contributions to the expectation values of $\mathcal O_\psi$ in \cite{Ross:2009ar}.
\end{itemize}
We see that the solutions of the linearized equations of motion divide into source terms and the corresponding response functions, as expected. Note that because of the conservation equations and the conformal constraint relating energy density to spatial stress, the number of  independent response functions and the number of sources does not match up. This could be resolved by working with gauge-independent combinations of the sources.\footnote{An error in the first version of this paper has been corrected here in light of the discussion in \cite{BBH}.}

One might have expected that we should have a vector operator corresponding to the field $A_\mu$, but this is not correct: because we can choose $A_I = 0$ by a choice of frame, there is only one piece of boundary data associated with this field, $A_0$, so there is a single scalar operator dual to changes in this boundary data. Physically, what happened is the spatial vector worth of additional boundary data in $A_\mu$ was absorbed by the frame fields, providing the extra information needed to have independent sources for $\mathcal E^i$ and $\mathcal P_i$. 

There are a few interesting open issues already at the level of the linearised analysis. In \cite{Ross:2009ar}, the prescription of section \ref{cft} was shown to give finite expectation values for all components of the stress tensor complex, once we added appropriate counterterms to the action. Surprisingly, there were divergences in $\mathcal E^i$ even when we set the modes associated with changes in the boundary data to zero, which required some derivative counterterms. In section \ref{ct}, we will see that the counterterms for asymptotically locally Lifshitz boundary conditions are uniquely fixed by cancelling the divergences arising from the boundary data. These counterterms should then also cancel the divergences involving the response functions; we comment on this issue in the discussion. 

There was also an issue with a divergence in the expectation value of $\mathcal O_\psi$ for $z \geq 2$. We will not investigate this further here, as this operator is irrelevant for $z \geq 2$, so considering sources for it would take us outside the asymptotically locally Lifshitz boundary conditions we are considering in this paper. The analysis could simply be extended by considering this source perturbatively, but we leave this for future work. However, we should note that there is an additional issue here which does not appear for example for irrelevant scalars in AdS: because of the coupled structure of the equations, the source mode $c_3$ appears linearly in the frame fields as well as in the vector field. Thus, we cannot think of varying $c_3$ as simply a variation of $A_0$, even perturbatively; it necessarily involves a variation of the frame fields as well. 

The most interesting open issue, however, is the crossover between the source mode $c_{4i}$ and the expectation value $c_{2i}$ in $v_{2i}$. When $z > 4$, the linearized mode with coefficient $c_{2i}$ falls off slower at large $r$ than the $c_{4i}$ mode. As a result, turning on non-zero $c_{2i}$ will violate our asymptotically locally Lifshitz boundary conditions. That is, imposing asymptotically locally Lifshitz boundary conditions for $z > 4$ gives us a smaller space of solutions than expected in the asymptotic regime; this will generically lead to problems satisfying the regularity conditions in the interior of the spacetime for arbitrary sources. We are therefore motivated to find different boundary conditions in this regime. At the linearized level, it is easy to see what we should do: considering the mixed boundary condition 
\begin{equation}
\lim_{r \to \infty} r^{-1} (\dot{e}^{(I)}_t - (z-3) e^{(I)}_t) = (4-z) v^I(x^\alpha), 
\end{equation}
$v^I(x^\alpha)$ would provide a source for the momentum density $\mathcal P_i$ while leaving the mode with $e^{(I)}_t \sim r^{z-3}$ free. Or we could continue to impose a Dirichlet boundary condition, but for $z > 4$ relax this to require that at large $r$,
\begin{equation}
e^{(I)}_t \sim r^{z-3} w^I(x^\alpha).
\end{equation}
The mode with $e^{(I)}_t \sim r$ is then free. This should correspond to a Legendre-transformed theory where we regard $\mathcal P_i$ as a source current, coupled to a vector operator of dimension $z-1$. The value of $w^I(x^\alpha)$ then fixes the value of $\mathcal P_i$, thought of as the source. The Dirichlet boundary condition fixes the leading term in $e^{(I)}_t$, so it is the analogue of the usual quantization for a scalar, while the mixed boundary condition fixing the source for $\mathcal P^i$ is the analogue of the alternative quantization. As noted in the previous section, for $z > 4$ the operator  $\mathcal P_i \mathcal P^i$ generates a flow from the fixed source boundary condition in the UV, which we expect to run to fixed $\mathcal P^i$ boundary conditions in the IR. Hence the latter are more generic in this range. Extending this analysis beyond the linearised level is an important open question, but we will leave it for future work and focus on the asymptotically locally Lifshitz boundary conditions \eqref{frame} in this paper.  

\section{Asymptotic expansion}
\label{eom}

In this section, we want to go beyond the linearised analysis by showing that solutions of the bulk equations of motion exist for arbitrary boundary data. To do so, we will solve the equations of motion in an asymptotic expansion: that is, we work at large $r$, and solve the equations in an expansion in powers of $r$. To treat this large $r$ expansion, it is convenient to rewrite the equations of motion in a radial Hamiltonian framework. If we work in Gaussian normal coordinates, the canonical coordinates on a surface $r=  r_0$ are the induced metric $h_{\alpha \beta}$ and the gauge field $A_\alpha$. The momenta are $\pi_{\alpha \beta} = K_{\alpha \beta} - h_{\alpha \beta} K$ and $\pi_\alpha = n^\mu F_{\mu \alpha} = r F_{r \alpha}$, where $K_{\alpha \beta}$ is the extrinsic curvature of the surface.\footnote{Canonical momenta would include a factor of $\sqrt{-h}$, but this convention will simplify the relation to the vevs.}

To make contact with the previous section, we may note that the momenta $\pi_{\alpha \beta}$, $\pi_\alpha$ can be related to the expectation values of the operators in the dual field theory by observing that on-shell, the variation of the action \eqref{Dact} is 
\begin{equation} \label{deltaS}
\delta S = \int d^3 x \sqrt{-h} (\pi^{\alpha \beta} \delta h_{\alpha \beta} + \pi^\alpha \delta A_\alpha).
\end{equation}
When we do frame variations $\delta e^{(A)}_\alpha$ holding $A_A$ fixed, the variation of the action on-shell is then
\begin{equation} \label{dS}
\delta S = \int d^3 x \sqrt{-h} ( (2 \pi^{\alpha}_{\ \beta} + \pi^\alpha A_\beta) e^\beta_{(B)} \delta e^{(B)}_\alpha  + \pi^A \delta A_A),
\end{equation}
where $\pi^A = \pi^\alpha e_\alpha^{(A)}$. Thus, we can define an object related to the boundary stress tensor complex considered in the previous section:
\begin{equation} \label{tstress}
T_{\alpha B}^{(bare)} = (2 \pi_{\alpha \beta} + \pi_\alpha A_\beta) e^\beta_{(B)}.
\end{equation}
Note that the variations in \eqref{dS} are with respect to bulk fields, whereas the variations in \eqref{fvar} are with respect to the boundary data. Thus this bulk object will differ from the boundary stress tensor by some powers of $r$. We also call this the ``bare'' stress tensor because using the naive Dirichlet action \eqref{Dact} gives a divergent stress tensor which will require renormalization. We address this in the next section. The point at this stage is just to note the relation between the momenta and the stress tensor complex. Unlike in the AdS case, we cannot simply trade the stress tensor for $\pi_{\alpha\beta}$; the $T_{\alpha B}$ also include information from $\pi_\alpha$.

\subsection{Gauss-Codazzi equations}

To solve the equations of motion in an expansion in powers of $r$, it is useful to decompose the equations in the radial direction, using the Gauss-Codazzi equations to rewrite them as a pair of dynamical equations,
\begin{align} \label{Kd}
\dot{K}_{\alpha \beta} + K K_{\alpha \beta} - 2 K_{\alpha \gamma} K^\gamma_{\ \beta} =& R_{\alpha \beta} - \Lambda h_{\alpha \beta} - \frac{1}{2} F_{\alpha \gamma} F_\beta^\gamma + \frac{1}{8} h_{\alpha \beta} F_{\gamma \sigma} F^{\gamma \sigma} - \frac{1}{2} \pi_\alpha \pi_\beta \nonumber \\ &+ \frac{1}{4} h_{\alpha \beta} \pi_\gamma \pi^\gamma - \frac{1}{2} m^2 A_\alpha A_\beta, 
\end{align}
\begin{equation} \label{pd}
\dot \pi^\alpha + K \pi^\alpha + \nabla_\beta F^{\beta \alpha} = m^2 A^\alpha, 
\end{equation}
and a set of constraints, 
\begin{equation} \label{dK}
\nabla_\alpha K^\alpha_{\ \beta} - \nabla_\beta K^\alpha_{\ \alpha} = \frac{1}{2} F_{\beta \alpha} \pi^\alpha + \frac{1}{2} m^2 A_\beta r A_r, 
\end{equation}
\begin{equation} \label{Ks}
K^2 - K_{\alpha \beta} K^{\alpha \beta} = R - 2 \Lambda + \frac{1}{2} \pi_\alpha \pi^\alpha - \frac{1}{4} F_{\alpha \beta} F^{\alpha \beta} + \frac{1}{2} m^2 r^2 A_r^2 - \frac{1}{2} m^2 A_\alpha A^\alpha, 
\end{equation}
and
\begin{equation} \label{Arr}
\nabla_\alpha \pi^\alpha = - m^2 r A_r. 
\end{equation}
Here $\dot{}$ denotes $r \partial_r$, and the Ricci tensor $R_{\alpha \beta}$ and covariant derivatives $\nabla_\alpha$ are with respect to the induced metric $h_{\alpha \beta}$ on a surface of constant $r$. Using \eqref{tstress}  and \eqref{Arr}, we can see that the constraint equation \eqref{dK} encodes the conservation equation \eqref{cons}. This is as expected; this is the diffeomorphism constraint in the radial Hamiltonian framework, associated with coordinate transformations on the surfaces of constant $r$. 

With our boundary conditions, the constraints \eqref{dK} and \eqref{Ks} are automatically satisfied when the dynamical equations are. This is because in general, the dynamical equations imply the $r$-derivative of the constraints, so the only non-trivial additional information in the constraints is the $r$-independent term. With asymptotically locally Lifshitz boundary conditions, the $r$-independent term reduces to the constraint in a Lifshitz spacetime, which is satisfied because this is a solution of the equations of motion. Thus, we need only consider the dynamical equations \eqref{Kd} and \eqref{pd}, substituting \eqref{Arr} for $r A_r$.

Since the asymptotic boundary condition is written in terms of frame fields, it will be convenient to also rewrite these equations in terms of frame fields $e^{(A)}$. As in the linearized analysis, we partially fix the choice of frame by setting $A_I = 0$, so that the distinguished timelike frame field is always aligned with the boundary component of the massive vector field. Note that the massive vector  field will still have a non-zero radial component, but this is determined algebraically by \eqref{Arr}. The relevant data is then the frame $e^{(A)}_\alpha$ and the matter field $A_0$.

For convenience, we define a ``frame extrinsic curvature'' 
\begin{equation}
K^A_{\ B} = e^\alpha_{(B)} \dot e^{(A)}_\alpha.
\end{equation}
Note that this is not a symmetric object, and the usual extrinsic curvature is $K_{\alpha \beta} = e^{(A)}_\alpha e^{(B)}_\beta K_{(AB)}$, where the round brackets denote symmetrisation. The equations of motion written with frame indices are then
\begin{align} \label{Kdot}
\dot K_{(AB)} + K K_{(AB)} + \frac{1}{2} (K_{CA} K^C_{\ B} - K_{AC} K_B^{\ C}) =& R_{AB} - \Lambda \eta_{AB} - \frac{1}{2} F_{AC} F_B^{\ C} + \frac{1}{8} \eta_{AB} F_{CD} F^{CD}\nonumber \\ & -\frac{1}{2} \pi_A \pi_B + \frac{1}{4} \eta_{AB} \pi_C \pi^C, 
\end{align}
\begin{equation} \label{pdot}
\dot{\pi}^A + K \pi^A - K^A_{\ B} \pi^B + \nabla_B F^{BA} = m^2 A^A, 
\end{equation}
and the constraints
\begin{equation} \label{Ksq}
K^2 - K_{(AB)} K^{AB} - \frac{1}{2} \pi_A \pi^A = R - 2 \Lambda - \frac{1}{4} F_{AB} F^{AB} + \frac{1}{2} m^2 A_n^2 - \frac{1}{2} m^2 A_A A^A, 
\end{equation}
\begin{equation} \label{DK}
\nabla^A K_{(AB)} - \nabla_B K^A_A = \frac{1}{2} F_{BA} \pi^A + \frac{1}{2} m^2 A_B A_n,
\end{equation}
\begin{equation} \label{Ar}
\nabla^A \pi_A = -m^2 A_n. 
\end{equation}
Here $A_n$ is the normal component of $A$, $A_n = n^\mu A_\mu = r A_r$, $F_{AB} = e_{(A)}^\alpha e_{(B)}^\beta F_{\alpha \beta}$, and $\nabla_A = e_{(A)}^\alpha \nabla_\alpha$, where the covariant derivative $\nabla_\alpha$ is a total covariant derivative (covariant with respect to both local Lorentz transformations and diffeomorphisms), determined by requiring $\nabla_\alpha e^{(A)}_\beta = 0$. The connection $\omega_{\alpha BC}$ defined by 
\begin{equation}
\nabla_\alpha V_B =  \partial_\alpha V_B - \omega_{\alpha B}^{\ \ \ C} V_C 
\end{equation}
is then related to the Ricci rotation coefficients by \cite{ortin}
\begin{equation} \label{ccrel}
\omega_{ABC} = -\Omega_{ABC} +  \Omega_{ACB} + \Omega_{BCA},
\end{equation}
where as a reminder we define the Ricci rotation coefficients by $de^{(C)} = \Omega_{AB}^{\ \ \ C} e^{(A)} \wedge e^{(B)}$. This implies $\omega_{[AB]}^{\ \ \ C} = - \Omega_{AB}^{\ \ \ C}$, and $\omega^{C}_{\ CD} = 2\Omega_{CD}^{\ \ \ C}$. Note that $\Omega_{AB}^{\ \ \ C}$ is antisymmetric in its first two indices, but the orthonormality of the frame fields implies that $\omega_{ABC}$ is antisymmetric in its last two indices. 

\subsection{Asymptotic analysis}

We want to study the equations of motion in the asymptotic regime by making an expansion in powers of $r$. The aim in this section is to analyse the structure of the equations of motion and demonstrate that a solution exists. We will find that asymptotically locally Lifshitz solutions exist for arbitrary values of the sources for $z < 2$, and if we set some of the sources to zero solutions continue to exist for $z \geq 2$. We first consider $z <2$ in detail, and then comment on the differences for $z \geq 2$. 

We recall that $e^{(0)} = r^z \hat{e}^{(0)}$ and $e^{(I)} = r \hat{e}^{(I)}$, and write $A_0 = \alpha + \psi$. The leading term in $\hat{e}^{(A)}$ is independent of $r$, and gives the boundary data dual to the stress tensor. From the linearized analysis, we see that $\psi = r^{-\Delta_- } \hat{\psi}(r,x^\alpha)$, where $\Delta_- = z+2 - \Delta_\psi = \frac{1}{2} (z+2 - \beta_z)$. The asymptotic value $\lim_{r \to \infty} \hat \psi$ corresponds to the source for the dual operator $\mathcal O_\psi$. For $z <2$, this is a relevant operator, so $\Delta_-$ is positive. We want to show that having chosen these leading terms, we can iteratively find a solution of \eqref{Kd}, \eqref{pd} in an expansion in powers of $r$ by adding subleading terms to $\hat{e}^{(A)}$ and $\hat \psi$. 

It might appear that we have fewer equations than unknowns, since we have nine components $ \hat{e}^{(A)}_\alpha$ plus one $\hat \psi$, and only nine equations. This is because there is a residual freedom to make a local rotation in the choice of the spatial frame vectors $e^{(I)}$, giving us one pure gauge mode.\footnote{In general dimension, there are $d^2+1$ fields, but the spatial rotations make $\frac{1}{2} d_s (d_s-1) = \frac{1}{2}(d-1)(d-2)$ of them pure gauge, leaving $\frac{1}{2} d(d+1)+d$ physical fields, matching the number of equations.} It is convenient to fix this gauge symmetry by choosing the spatial frames so that $K_{IJ}$ is symmetric. 

Although $\hat{e}^{(A)}$ and $\hat \psi$ are the fundamental quantities we are solving for, in analysing \eqref{Kd} and \eqref{pd} it is more convenient to work in terms of $K_{AB}$ and $\pi^A$. These are given by 
\begin{align}
K_{00} =&\, -z + \dot{\hat e}_{(0) \alpha} \hat e_{(0)}^\alpha, \\
K_{0I} =&\, r^{z-1} \dot{\hat e}_{(0) \alpha} \hat e_{(I)}^\alpha, \nonumber \\
K_{I0} =&\, r^{1-z} \dot{\hat e}_{(I) \alpha} \hat e_{(0)}^\alpha, \nonumber \\
K_{IJ} =&\, \delta_{IJ} + \dot{\hat e}_{(I) \alpha} \hat e_{(J)}^\alpha, \nonumber
\end{align}
which implies $K = z+2+ \dot{\hat e}_{(A) \alpha} \hat e^{(A) \alpha}$, and 
\begin{align}
\pi_0 =&\, \dot \psi -K_{00} (\alpha + \psi) - \partial_0 A_n, \\
\pi_I =&\, -K_{0I} (\alpha + \psi) - \partial_I A_n. \nonumber
\end{align}
Working in an expansion in powers of $r$, we can solve \eqref{Ar} at each order to determine $A_n$. Because of the derivative, this will be determined in terms of contributions at earlier orders.

There is an explicit positive power in $K_{0I}$; to have a solution where all quantities involve only negative powers, the dependence of  $\dot{\hat e}_{(0) \alpha} \hat e_{(I)}^\alpha$ must be sufficiently suppressed so that the expansion of $K_{0I}$ contains only negative powers. We will see that we can self-consistently find such a solution. 

The curvature and field strength terms will also involve explicit powers. The key point is to see that these all involve negative powers, so that the sources produced by considering arbitrary boundary data can be cancelled by adding subleading terms, and the corrections produced by the curvature of the subleading terms are further subleading. This turns out to be true for the Ricci rotation coefficients:
\begin{align}
\Omega_{0I}^{\ \ 0} = &\, r^{-1} (d \hat e^{(0)})_{\alpha \beta}  \hat e^{[\alpha}_{(0)}\hat e^{\beta]}_{(I)} , \\
\Omega_{IJ}^{\ \ \ K} = &\, r^{-1} (d \hat e^{(K)})_{\alpha \beta}  \hat e^{[\alpha}_{(I)}\hat e^{\beta]}_{(J)} , \nonumber \\
\Omega_{0I}^{\ \ J} = &\, r^{-z} (d \hat e^{(J)})_{\alpha \beta}  \hat e^{[\alpha}_{(0)}\hat e^{\beta]}_{(I)} , \nonumber \\
\Omega_{IJ}^{\ \ \ 0} = &\, r^{z-2} (d \hat e^{(0)})_{\alpha \beta}  \hat e^{[\alpha}_{(I)}\hat e^{\beta]}_{(J)} , \nonumber
\end{align}
Explicit derivatives will also come with similar factors: $\partial_0 = e_{(0)}^\alpha \partial_\alpha = r^{-z} \hat e_{(0)}^\alpha \partial_\alpha$ and $\partial_I = e_{(I)}^\alpha \partial_\alpha = r^{-1} \hat e_{(I)}^\alpha \partial_\alpha$.

We might have expected some positive powers to appear, since we are considering turning on sources for the irrelevant operator $\mathcal E^i$, and the breakdown of the UV behaviour associated with irrelevant operators is normally signalled in the bulk through the appearance of growing terms in these equations (see \cite{vanRees:2011fr} for a recent discussion in the AdS case). The term that comes closest to this expectation is $\Omega_{IJ}^{\ \ \ 0}$, but it has an extra factor of $r^{-1}$ because it involves the derivatives of the source for this operator, rather than the source itself. A constant value of this source corresponds to a diffeomorphism, so only its derivatives will enter into the equations of motion, and thus for $z <2$ it will not lead to positive powers in the expansion despite the operator being irrelevant.  There is still a change in the UV behaviour when we have non-zero sources for $\mathcal E^i$, but it is signalled only by the violation of \eqref{irr}, implying that the boundary does not have a notion of absolute time.

Using \eqref{ccrel}, we can express the curvature terms in the field strength and the Ricci tensor in terms of the Ricci rotation coefficients. For the field strength,
\begin{equation}
F_{AB} = \partial_A A_B - \partial_B A_A - \omega_{AB}^{\ \ \ C} A_C + \omega_{BA}^{\ \ \ C} A_C = \partial_A A_B - \partial_B A_A + 2 \Omega_{AB}^{\ \ \ C} A_C.
\end{equation}
Hence $F_{0I} = -\partial_I A_0 + 2 \Omega_{0I}^{\ \ 0}A_0$ will contain an explicit $r^{-1}$, and $F_{IJ} = 2 \Omega_{IJ}^{\ \ \ 0}A_0$ will contain an explicit $r^{z-2}$. The covariant derivative term appearing in \eqref{pdot} is
\begin{equation}
\nabla^A F_{AB} = \partial^2 A_B - \partial^A \partial_B A_A + 2 \partial^A (\Omega_{AB}^{\ \ \ C} A_C) - 2 \Omega_C^{\ DC} F_{DB} - \Omega^{AD}_{\ \ \ B} F_{AD},
\end{equation}
so $\nabla^A F_{A0}$ will have terms containing $r^{-2}$ and a term quadratic in $\Omega_{IJ}^{\ \ \ 0}$ containing $r^{2z-4}$, while $\nabla^A F_{AI}$ will have terms containing $r^{-(1+z)}$ and a term linear in $\Omega_{IJ}^{\ \ \ 0}$ containing $r^{z-3}$. The Ricci tensor is 
\begin{align}
R_{AB} =& - \partial_A \omega_{C \ B}^{\ C} - \partial_C \omega_{AB}^{\ \ \ C} + \omega_{CDA} \omega^{DC}_{\ \ \ B} + \omega_{ABD} \omega_{C}^{\ CD}, \\
=& -2 \partial_A \Omega_{CB}^{\ \ \ C} + \partial_C ( \Omega_{AB}^{\ \ \ C} - \Omega_{B\ \ A}^{\ \ C} - \Omega_{A\ \ B}^{\ \ C}) - \Omega_{CDA} \Omega^{CD}_{\ \ \ B} + 4\Omega_{CAD} \Omega^{(C\ \ D)}_{\ \ B} \nonumber \\ &+ 2 \Omega_C^{\ DC} (\Omega_{BAD} + \Omega_{BDA} + \Omega_{ADB}). \nonumber
\end{align}
As a result, $R_{00}$ and $R_{IJ}$ will contain terms with $r^{-2z}$, $r^{-2}$, and a term quadratic in $\Omega_{IJ}^{\ \ \ 0}$ containing $r^{2z-4}$, while $R_{0I}$ will contain terms with   $r^{-(1+z)}$ and a term linear in $\Omega_{IJ}^{\ \ \ 0}$ containing $r^{(z-3)}$.

Thus, all the explicit powers are negative, and we can construct a solution by making an expansion in negative powers.  Let us write 
\begin{equation}
K_{AB} = \sum_\Delta r^{-\Delta} K_{AB}^{(\Delta)}, \quad \psi = \sum_\Delta r^{-\Delta} \psi^{(\Delta)}.
\end{equation}
Assuming the sum only involves negative powers, that is $\Delta  \geq 0$, then the terms in \eqref{Kd}, \eqref{pd} of order $r^{-\Delta_0}$ will involve $K_{AB}^{(\Delta)}$, $\psi^{(\Delta)}$ for $\Delta \leq \Delta_0$, and curvature terms. The $00$ and $IJ$ components of \eqref{Kd} and the $0$ component of \eqref{pd} can be solved to determine the $\Delta = \Delta_0$ components of $K_{00}$, $K_{IJ}$ and $\psi$, while the $0I$ components of \eqref{Kd} and the $I$ components of \eqref{pd} determine the $\Delta = \Delta_0$ components of $K_{0I}$ and  $K_{I0}$. The sources are given by the quadratic terms involving the extrinsic curvature terms and $\pi_A$ at earlier orders, and the curvature terms which involve the frame fields and $\psi$ at earlier orders. 

To determine which powers occur, we work iteratively starting from the leading source terms coming from derivatives of the boundary data. The curvature of the $r$-independent terms in $\hat{e}^{(A)}_\alpha$ will give sources in the $00$ and $IJ$ components of \eqref{Kdot} and the $0$ component of \eqref{pd}. These sources go like $r^{-2z}$, $r^{-2}$, and $r^{2z-4}$. Similarly substituting the curvature of the $r$-independent background values will produce terms in the $0I$ components of \eqref{Kdot} and the $I$ components of \eqref{pdot}  which go like $r^{-(1+z)}$ and $r^{z-3}$. To cancel all of these terms, we need subleading terms in $K_{AB}$, $\psi$ with these same falloffs.  Thus, we would need subleading terms in the frame fields such that
\begin{align} 
\dot{\hat e}_{(0) \alpha} \hat e_{(0)}^\alpha \sim r^{2z-4},  r^{-2}, r^{-2z},  \\
\dot{\hat e}_{(I) \alpha} \hat e_{(J)}^\alpha \sim r^{2z-4},  r^{-2}, r^{-2z},  \nonumber \\ \dot{\hat e}_{(0) \alpha} \hat e_{(I)}^\alpha \sim r^{-2}, r^{-2z}, \nonumber \\
\dot{\hat e}_{(I) \alpha} \hat e_{(0)}^\alpha \sim r^{2z-4},  r^{-2}, r^{-2z},  \nonumber
\end{align}
Note that because of the explicit $r^{z-1}$ in $K_{0I}$, the required behaviour for $\dot{\hat e}_{(0) \alpha} \hat e_{(I)}^\alpha$ does not involve the leading $r^{2z-4}$ falloff that appears in all the other terms. That is, the presence of the explicit positive power in $K_{0I}$ has the effect that the behaviour of the subleading terms in $\hat e_{(0) \alpha}$ obtained by solving these equations will produce a faster falloff for the term contributing to $K_{0I}$.

If we include the effect of the leading term in $\psi$, this will give contributions to the equations with powers of $r^{-\Delta_-}$. This can appear on its own in the $00$ component of \eqref{Kdot} and in the $0$ component of \eqref{pdot}, and in combination with derivative terms in all the equations. Since $A_A$ appears quadratically in \eqref{Kdot}, there will also be terms going like $r^{-2\Delta_- }$. To cancel these contributions, we will need to add additional subleading contributions containing powers $r^{-\Delta_- }$. These are negative powers for $z < 2$. 

All of these subleading terms will then in turn contribute to the curvature terms and to the quadratic terms in \eqref{Kd}, \eqref{pd} requiring further subleading terms to cancel these contributions. We will thus need an infinite series of contributions to $\hat{e}^{(A)}$, $\hat \psi$, with powers
\begin{equation} \label{Delta}
\Delta = 2m + 2nz + (2z-4) p + \Delta_- q, \quad m, n, p, q \in 0,1,2, \ldots
\end{equation}

We have concentrated above on subleading terms in the large $r$ expansion which are locally determined by the boundary data. However, there will also be independent terms in the expansion which are not locally determined by the boundary data, corresponding to the second set of linearised solutions in section \ref{lin}. These are referred to as the response functions, as they determine the finite part of the expectation values, as can be seen from the linearised analysis of \cite{Ross:2009ar}.  These appear first at specific characteristic powers of $r$ corresponding to the dual operator dimensions. The leading-order terms will be given precisely by the linearised solution in the previous section, as the $x^\alpha$ dependence of the boundary data does not effect the radial equation for the leading-order part of these modes. 

In the asymptotic expansion, we can see the leading $r$-dependence of these additional contributions  by finding the value of $\Delta$ where the equations at order $\Delta$ have a non-trivial solution with no source contribution. This is most easily illustrated by considering the trace-free part of the $IJ$ components in \eqref{Kd}, for which the leading contribution to the equations at $e^{-\Delta r}$ is simply $-\Delta K_{IJ}^{(\Delta)} + (z+2) K_{IJ}^{(\Delta)}$, so that in the absence of sources there is a solution at $\Delta = z+2$, corresponding to the modes $t_{d2}$ and $t_{o2}$ in section \ref{lin}. This works similarly for the other modes. but the coupling between the equations makes the analysis more complicated. For generic $z$, these response functions appear at a value of $\Delta$ where there is no source contribution. For some $z$, the $\Delta$ determined in this way is one of the values \eqref{Delta}, and then there will be a logarithmic solution; we will ignore these cases for simplicity, but it should be easy to include them. The curvature terms arising from the leading part of the response functions will source another infinite series of subleading corrections. 

Thus in summary, for $z < 2$, there are solutions of the bulk equations of motion for arbitrary boundary data, which can be constructed order by order in an expansion in powers of $r$. There is an infinite series of terms with powers \eqref{Delta} which are determined by the boundary data. The solution in this asymptotic expansion is not unique, as there are additional response functions in the general solution which are not determined locally in terms of the boundary data. One expects these to be fixed in terms of the boundary data by imposing appropriate regularity or inner boundary conditions in the interior of the spacetime. In our analysis of the counterterms in the next section we will assume this is true, so that the bulk solution is uniquely determined by the boundary data, but we will not explore it in detail, as we are restricting to an asymptotic analysis.

\subsubsection{Expansion for $z >2$}

In the above analysis, we assumed that $z < 2$. If $z \geq 2$, then $\Delta_- < 0$, and to have an asymptotically locally Lifshitz solution, we need to set the source term $\lim_{r \to \infty} \hat \psi$ to zero. This is a familiar story from the AdS context; adding a source for an irrelevant operator will change the UV behaviour of the theory, which is reflected in the bulk dual by a violation of the usual asymptotic boundary conditions.

Additionally, the explicit power appearing in $\Omega_{IJ}^{\ \ \ 0}$ is now positive, so in order to have a solution with all negative powers, we have to restrict the leading contribution to $\hat{e}^{(0)}_\alpha$ such that it is irrotational. That is, we now need to set the source for the energy flux $\mathcal E^i$ to zero to have asymptotically locally Lifshitz solutions satisfying \eqref{frame}. 

With this restriction, the leading contributions in  the $00$ and $IJ$ components of \eqref{Kdot} and the $0$ component of \eqref{pd} from the curvature of the boundary data will go like $r^{-2z}$ and $r^{-2}$, and the leading contributions in the $0I$ components of \eqref{Kdot} and the $I$ components of \eqref{pdot} from curvature of the boundary data will go like $r^{-(1+z)}$. To cancel these, we need subleading terms in $K_{AB}$, $\psi$ with these same falloffs.  Thus, we would need subleading terms in the frame fields such that
\begin{align} 
\dot{\hat e}_{(0) \alpha} \hat e_{(0)}^\alpha \sim  r^{-2}, r^{-2z},  \\
\dot{\hat e}_{(I) \alpha} \hat e_{(J)}^\alpha \sim  r^{-2}, r^{-2z},  \nonumber \\ \dot{\hat e}_{(0) \alpha} \hat e_{(I)}^\alpha \sim r^{-2z}, \nonumber \\
\dot{\hat e}_{(I) \alpha} \hat e_{(0)}^\alpha \sim  r^{-2}, r^{-2z},  \nonumber
\end{align}
Again, we see a faster falloff in $\dot{\hat e}_{(0) \alpha} \hat e_{(I)}^\alpha$. As before, this faster falloff ensures that $K_{0I}$ will involve the same powers as $K_{I0}$, despite the positive power in the expression for $K_{0I}$. It also implies that  the first non-zero term in $\Omega_{IJ}^{\ \ \ 0}$ comes from the terms at $\Delta = 2z$, and hence goes like $r^{-4}$. 

Thus, for $z >2$, once we fix the sources for $\mathcal O_\psi$ and $\mathcal E^i$ to be zero, the powers appearing in the expansion of  \eqref{Kdot} and \eqref{pd} will start with $r^{-2}$ and $r^{-2z}$, and for arbitrary values of the other boundary data, there will be a solution in a series of negative powers as before, with powers
\begin{equation}
\Delta = 2m + 2nz, \quad m, n \in \mathbb{N}, 
\end{equation}
 together with the response function contributions from section \ref{lin} and their subleading corrections. 

Note however that as noted in section \ref{lin}, if we impose asymptotically locally Lifshitz boundary conditions for $z \geq 4$, the size of the space of response functions is reduced,  because the mode $c_{2i}$ has to be set to zero to satisfy the boundary conditions. We therefore expect that we will not generically have enough freedom to satisfy the inner boundary condition for arbitrary boundary data. That is, for $z \geq 4$, while we will have asymptotically locally Lifshitz solutions in the asymptotic regime, to have an appropriate space of regular solutions for arbitrary boundary data we will need to consider a different boundary condition as outlined in section \ref{lin}. 
 
\section{Holographic renormalization}
\label{ct}

Having explored the space of solutions for our boundary conditions, we should now construct an appropriate action principle for arbitrary boundary data to complete the holographic dictionary. To be able to apply the prescription given in section \ref{cft}, we need an action which is finite on-shell and has vanishing variation on-shell under arbitrary variations satisfying the asymptotic boundary conditions. That is, $\delta S = 0$ on-shell if the field variation does not change the boundary data. We will construct such an action by adding  appropriate local boundary counterterms to cancel the divergences in the action \eqref{Dact}. 

We will give an efficient algorithm for obtaining the required counterterms, based on the approach of  \cite{Papadimitriou:2004ap,Papadimitriou:2004rz}. This involves expanding in eigenvalues of an appropriate bulk dilatation operator, rather than in powers of $r$ as we did in the previous section. The first step is to take a Hamilton-Jacobi style approach to determining the on-shell action. Assuming that we impose some appropriate boundary or regularity condition in the interior of the spacetime, the on-shell solution of the equations of motion will be uniquely determined in terms of the asymptotic boundary data, so the on-shell action is a function of the boundary data, which we can write as a boundary term,
\begin{equation} \label{Slambda}
S  =  \int d^3x \sqrt{-h} \lambda(e^{(A)}, \psi).
\end{equation}
Note that we actually write the action as a function of the values of the bulk fields evaluated on the boundary, rather than in terms of the boundary data. For asymptotically locally Lifshitz boundary conditions, the leading asymptotic values of the fields are determined by the boundary data, so this contains equivalent information, but it is better to work in terms of the boundary values of bulk fields, since counterterms added to the action need to be written in terms of the latter.
 
The variation of the on-shell action is given by \eqref{deltaS}, and we can use this to replace the dynamical ODEs \eqref{Kdot}, \eqref{pdot}  by functional differential equations, writing 
\begin{equation} 
\pi^{\alpha \beta} =  \frac{1}{\sqrt{-h}} \frac{\delta}{\delta h_{\alpha \beta}}  S, \quad \pi^\alpha =\frac{1}{\sqrt{-h}} \frac{ \delta}{\delta A_\alpha} S.
\end{equation}
These relations give the canonical momenta as functions of the boundary data. Assuming that the on-shell action is boundary diffeomorphism invariant will imply that these quantities automatically satisfy the diffeomorphism constraint \eqref{dK}, and if we substitute them into the Hamiltonian constraint \eqref{Ks}, this becomes a quadratic equation in the functional derivatives of $\lambda$, which can be solved to determine $\lambda$ in an appropriate expansion. The divergences we need to cancel are determined by applying this procedure to the naive action \eqref{Dact}, calculating its on-shell value as a function of the boundary data. Since we are working with divergent quantities, we will need to introduce a regulator. We therefore assume that the boundary integral in \eqref{Slambda} is evaluated on a surface $r=r_0$, and the value of the on-shell action is then a functional of the values of the bulk fields evaluated on this surface; these Dirichlet data determine the value of the action for the bulk solution. Also note that the value of the on-shell action is in general a non-local function of the boundary data, as it involves the response functions, which are only determined once we impose regularity conditions in the interior of the spacetime. An essential point is to see that the divergent terms in the naive on-shell action \eqref{Dact} are local functions of the boundary data, and do not involve the response functions, so they can be cancelled by simply subtracting the same function as a boundary counterterm in our definition of the action. 

For the Lifshitz case we need to introduce slightly different variables, defining the functional derivatives
\begin{equation} \label{Pi}
T^A_{\ B} =  \frac{1}{\sqrt{-h}} e_\alpha^{(A)}\frac{ \delta}{\delta e^{(B)}_\alpha} S = \frac{1}{\sqrt{-h}} e_\alpha^{(A)}\frac{ \delta}{\delta e^{(B)}_\alpha} \frac{1}{16 \pi G_4} \int d^3x \sqrt{-h} \lambda
\end{equation}
and
\begin{equation} \label{p}
\pi_\psi = \frac{1}{\sqrt{-h}} \frac{\delta}{\delta \psi} S = \frac{1}{\sqrt{-h}} \frac{\delta}{\delta \psi} \frac{1}{16 \pi G_4} \int d^3 x \sqrt{-h} \lambda.
\end{equation}
This change of variables is important because our boundary conditions give definite falloff behaviour for the frame fields, not the boundary metric.  From the field theory point of view, \eqref{Pi} and \eqref{p} are convenient variables because they will determine the vevs of the corresponding boundary operators (we explain the relation to the vevs in a little more detail below).  We will choose a gauge such that $T_{IJ}$ is symmetric, so that the number of quantities in \eqref{Pi} and \eqref{p} matches the number of equations we are replacing. From the relation \eqref{tstress}, we see that we can obtain the canonical momenta from these new variables, by writing $\pi^0 = \pi_\psi$,  using
\begin{equation} \label{pii}
\pi_I A_0 = T_{I0} - T_{0I}
\end{equation}
to determine $\pi_I$, and recovering $\pi_{AB}$ from  
\begin{equation} 
\pi_{AB} = \frac{1}{2} (T_{AB} - \pi_A A_B).  
\end{equation}
Thus, the canonical momenta are determined in terms of functional derivatives of the action, and the constraint \eqref{Ksq} can be solved in an appropriate expansion to determine the value of the on-shell action density $\lambda$. 

We want to introduce a functional derivative which agrees at leading order with the radial derivative, which will be used to organise the expansion of the on-shell action. Assuming our asymptotically locally Lifshitz boundary conditions \eqref{frame}, we have that $e^{(0)}_\alpha$ at leading order goes like $r^{z}$ and $e^{(I)}_\alpha$ at leading order goes like $r$, while from the linearized solution, at leading order $\psi$ goes like $r^{-\Delta_-}$. Thus, if we introduce the dilatation operator
\begin{equation} \label{dD}
\delta_D = -\int d^3 x \left( z e_\alpha^{(0)} \frac{ \delta}{\delta e_\alpha^{(0)}} + e_\alpha^{(I)} \frac{ \delta}{\delta e_\alpha^{(I)}} - (z+2-\Delta_\psi) \psi\frac{\delta}{\delta \psi} \right),
\end{equation}
then acting on any function of $e^{(A)}$, $\psi$, this will agree with the radial derivative to leading order at large $r$, $\delta_D \sim -r \partial_r$.\footnote{Note that, as stressed recently in \cite{vanRees:2011fr}, this statement depends crucially on satisfying our asymptotic boundary conditions, and it is difficult to extend this dilatation approach to consider even perturbative violations of the boundary conditions.} We can also see that $\delta_D$ can be viewed as a bulk analogue of the field theory dilatation operator; the field theory operator would have the same form as \eqref{dD}, but with the bulk fields replaced by boundary data. From \eqref{Pi} and \eqref{p}, we can see that if we consider the dilatation operator acting on the on-shell action, we can replace the functional derivatives on the RHS acting on the action by $T^A_{\ B}$, $\pi_\psi$. As a result, we have a relationship
\begin{equation} \label{lambda}
(z+2-\delta_D) \lambda = z T^0_{\ 0} + T^I_{\ I} - (z+2-\Delta_\psi) \psi \pi_\psi. 
\end{equation}

We now expand all these quantities in eigenfunctions of the dilatation operator $\delta_D$, writing
\begin{equation} \label{dilexp}
\lambda = \sum_{\Delta  \geq 0} \lambda^{(\Delta)}, \quad \delta_D \lambda^{(\Delta)} = \Delta \lambda^{(\Delta)}, 
\end{equation}
and similarly for $T^A_{\ B}$, $\pi_\psi$.  Note that because of the different scaling weights carried by the frame fields, the components of $T^A_{\ B}$ obtained by functionally differentiating $\lambda^{(\Delta)}$ do not all have the same eigenvalue $\Delta$:
\begin{equation} \label{ltoT}
\lambda^{(\Delta)} \to T^{0\  (\Delta)}_{\ 0}, T^{0\  (\Delta+1-z)}_{\ I}, T^{I\  (\Delta+z-1)}_{\ 0}, T^{I\  (\Delta)}_{\ J}.  
\end{equation}
Similarly, the fact that $\psi$ carries a non-trivial scaling weight implies 
\begin{equation} \label{ltopi}
\lambda^{(\Delta)} \to \pi_\psi^{(\Delta-\Delta_-)}.
\end{equation}
Expanding \eqref{lambda} in dilatation eigenvalues, we have
\begin{align} \label{lD}
(z+2-\Delta) \lambda^{(\Delta)} &= z T^{0\ (\Delta)}_{\ 0} + T^{I\ (\Delta)}_{\ I} - \Delta_- \psi \pi_\psi^{ (\Delta-\Delta_-)} \\ \nonumber
&= 2 z \pi^{0\ (\Delta)}_{\ 0} + 2 \pi^{I\ (\Delta)}_{\ I} + z \alpha \pi_\psi^{(\Delta)} + (z - \Delta_- )\psi \pi_\psi^{(\Delta-\Delta_-)}. 
\end{align}

Note that \eqref{lD} does not determine $\lambda^{(\Delta)}$ for $\Delta = z+2$. This is the leading term in the dilatation expansion associated with the response functions; it is also the term which makes a finite contribution to the on-shell action, as it scales like $r^{-(z+2)}$ at leading order, cancelling the scaling of the integration measure $\sqrt{-h}$.  From the field theory point of view, the expectation values of the operators $T_{AB}$ and $O_\psi$ are the functional derivatives of the on-shell bulk action with respect to the boundary data. Since the boundary data give the leading terms in the bulk fields, the functional derivatives with respect to boundary data can be exchanged for the ones involving $e^{(A)}$, $\psi$ in \eqref{Pi}, \eqref{p} up to some overall powers of $r$. Once we have cancelled the divergent contributions, the leading contribution to the action comes from $\lambda^{(z+2)}$, so the expectation values will be given by its functional derivatives, that is,
\begin{equation} 
\langle \mathcal E \rangle = r^{z+2} T^{0\  (z+2)}_{\ 0}, \quad \langle \mathcal E^i \rangle = r^{2z+1} T^{I\  (2z+1)}_{\ 0}, \quad \langle \mathcal P_i \rangle  = r^{3} T^{0\  (3)}_{\ I}, \quad \langle \Pi^i_{\ j} \rangle = r^{z+2}  T^{I\  (z+2)}_{\ J},
\end{equation}
and $\langle \mathcal O_\psi \rangle = r^{\Delta_\psi} \pi_\psi^{(\Delta_\psi)}$ (in all these quantities, the limit as $r \to \infty$ should be taken to compute the actual vevs). Note also that for $\Delta = z+2$ the right-hand side of \eqref{lD} must vanish. This is precisely the tracelessness condition for the stress tensor complex, generalised to the case where we explicitly break scale invariance by introducing  a source $\hat \psi$ for $\mathcal O_\psi$. 

Our interest is not however primarily in this undetermined part, but in the terms in the dilatation expansion of the on-shell action which are locally determined by the boundary data. These are determined by \eqref{lD} together with the constraint equation \eqref{Ksq}, whose expansion in dilatation eigenvalues is
\begin{align} \label{Hexp}
\sum_{s < \Delta/2}\left[ 2  K^{(s)} K^{(\Delta-s)} -2 K_{AB}^{(s)} K^{AB(\Delta-s)} - \pi_A^{(s)} \pi^{A(\Delta-s)} - \frac{1}{m^2} (\nabla_A \pi^A)^{(s)} (\nabla_A \pi^A)^{(\Delta-s)} \right] & \\ \nonumber  + \left[ \frac{1}{2} K^{(\Delta/2)2} -  K_{AB}^{(\Delta/2)} K^{AB(\Delta/2)} - \frac{1}{2} \pi_A^{(\Delta/2)} \pi^{A(\Delta/2)} - \frac{1}{2m^2} (\nabla_A \pi^A)^{(\Delta/2)2}  \right] &= src^{(\Delta)},
\end{align}
where $src^{(\Delta)}$ is the source contribution from the right-hand side of \eqref{Ksq}, which only appears for some particular values of $\Delta$. 

We want to manipulate this expression to obtain a formula for the right-hand side of \eqref{lD}. To do so, it is useful to note that the first two terms in \eqref{Hexp} can be rewritten as $2  K^{(s)} K^{(\Delta-s)} -2 K_{AB}^{(s)} K^{AB(\Delta-s)} = -2 K_{AB}^{(s)} \pi^{AB(\Delta-s)}$. Now from the above large $r$ expansion, we can see that 
\begin{equation} \label{K0}
K^{0\  (0)}_{\ 0} = z, \quad K^{I\  (0)}_{J} = \delta^I_{\ J}.
\end{equation}
For the vector momentum, we can write
\begin{equation}
\pi_A = \dot{A}_A + A_B K^B_{\ A} - \partial_A A_r, 
\end{equation}
which gives
\begin{equation}
\pi_0^{(0)} = \alpha K^{0\  (0)}_{\ 0} = z \alpha
\end{equation}
and a constraint on $\pi_0^{(\Delta_-)}$, 
\begin{equation} \label{pid}
\pi_0^{(\Delta_-)} = \dot{\psi} + \alpha K^{0\  (\Delta_-)}_{\ 0} + \psi K^{0\  (0)}_{\ 0} = \alpha K^{0\ (\Delta_-)}_{\ 0} + (z-\Delta_-) \psi. 
\end{equation}
Note that we can combine these results to obtain
\begin{equation} \label{Pi0}
T^{A\  (0)}_{\ B} = - 2(z+1)\delta^A_{\ B}. 
\end{equation}
For a generic value of $\Delta$, the terms in \eqref{Hexp} with $s=0$ and $s=\Delta_-$ are then 
\begin{align}
-2z \pi^{0\ (\Delta)}_{0} - 2 \pi^{I\ (\Delta)}_{I} - z \alpha \pi^{0(\Delta)} -2K_{AB}^{(\Delta_-)} \pi^{AB (\Delta - \Delta_-)} - \pi_A^{(\Delta_-)} \pi^{A (\Delta - \Delta_-)} \\  = - (z T^{0\ (\Delta)}_{0} +  T^{I\ (\Delta)}_{I}) - (\pi_0^{(\Delta_-)} - z \psi - \alpha K^{0\ (\Delta_-)}_{\ 0}) \pi^{0 (\Delta-\Delta_-)} \nonumber \\  -K_{AB}^{(\Delta_-)} T^{AB (\Delta - \Delta_-)} - \pi_I^{(\Delta_-)} \pi^{I (\Delta - \Delta_-)}+ \psi K^{0\ (\Delta_-)}_{\ 0}) \pi^{0 (\Delta -2 \Delta_-)}. \nonumber
\end{align}
Putting this together with \eqref{lD} gives
\begin{equation} \label{ldelta}
\boxed{(z+2-\Delta) \lambda^{(\Delta)} =  zT^{0\ (\Delta)}_{\ 0} + T^{I \ (\Delta)}_{I} - \Delta_- \psi \pi^{(\Delta-\Delta_-)}_\psi = - quad^{(\Delta)} - src^{(\Delta)}, }
\end{equation}
where $quad^{(\Delta)}$ is the remaining quadratic terms, 
\begin{align} \label{qexp}
quad^{(\Delta)} =& \sum_{0< s < \Delta/2 ; s \neq \Delta_-}\left[  2K_{AB}^{(s)} \pi^{AB(\Delta-s)} + \pi_A^{(s)} \pi^{A(\Delta-s)} + \frac{1}{m^2} (\nabla_A \pi^A)^{(s)} (\nabla_A \pi^A)^{(\Delta-s)} \right] \nonumber \\ &+  \left[ K_{AB}^{(\Delta_-)} T^{AB(\Delta-\Delta_-)} + K_{00}^{(\Delta_-)} \pi^{0(\Delta-2\Delta_-)} \psi  + \pi_I^{(\Delta_-)} \pi^{I(\Delta-\Delta_-)} \right] \nonumber \\ &+ \left[ K_{AB}^{(\Delta/2)} \pi^{AB(\Delta/2)} + \frac{1}{2} \pi_A^{(\Delta/2)} \pi^{A(\Delta/2)} + \frac{1}{2m^2} (\nabla_A \pi^A)^{(\Delta/2)2}  \right],
\end{align}
and the source terms will be given below. This is the equation we wish to solve to determine the divergent parts of the on-shell action. The advantage of this dilatation expansion, relative to the expansion in powers of $r$ we considered in the previous section, is that the curvature terms in \eqref{Ksq}, which involved an infinite series of terms in the expansion in powers of $r$, contribute a finite set of terms with definite dilatation eigenvalues, so we can write them all down explicitly, and the sub-leading terms contribute to further sub-leading terms only through the quadratic terms \eqref{qexp}. 

Let us first consider the source terms. For the Ricci rotation coefficients $\Omega_{AB}^{\ \ \ C}$, using the behaviour of the frame fields, the dilatation dimensions are $\Delta = 1$ for $\Omega_{0I}^{\ \ 0}$ and $\Omega_{IJ}^{\ \ \ K}$, $\Delta = z$ for $\Omega_{0I}^{\ \ J}$, and $\Delta = 2-z$ for $\Omega_{IJ}^{\ \ \ 0}$. For the vector field, $\psi$ has by definition dimension $\Delta = \Delta_-$. Hence $F_{0I} = -\partial_I A_0 + 2 \Omega_{0I}^{\ \ 0}A_0$ will contain components of dimension $1$, $1+\Delta_-$, and $F_{IJ} = 2 \Omega_{IJ}^{\ \ \ 0}A_0$ will contain components of dimension $2-z$, $2-z+\Delta_-$. The Ricci scalar in terms of the Ricci rotation coefficients is 
\begin{equation}
R = -4 \partial_A \Omega^{CA}_{\ \ \ C} + 4 \Omega^{CA}_{\ \ \ C} \Omega^{B}_{\ AB} + \Omega_{ABC} \Omega^{ABC} + 2 \Omega_{ABC} \Omega^{CBA}. 
\end{equation}
This will have contributions at dimensions $2$, $2z$ and $4-2z$. 

Putting all this together, we can obtain the source terms at each order:
\begin{itemize}
\item $\Delta = 0$:
\begin{equation} \label{s0}
src^{(\Delta)} = -2 \Lambda + \frac{1}{2} m^2 \alpha^2 = 2(z+1)(z+2). 
\end{equation}
\item $\Delta = 2$: 
\begin{eqnarray} \label{s2}
src^{(\Delta)} &=& R^{(2)} - \frac{1}{4} F_{AB}^{(1)} F^{AB(1)} = - 4 \partial_I \Omega^{CI}_{\ \ \ C} + 4 \Omega^{CI}_{\ \ \ C} \Omega^{B}_{\ \  IB} +4 \Omega_{0I0} \Omega^{0I0} \\ &&+ \Omega_{IJK} \Omega^{IJK} + 2 \Omega_{IJK} \Omega^{KJI} + 4 \Omega_{0JK} \Omega^{KJ0} + \Omega_{0I0}\Omega^{0I0} \alpha^2. \nonumber
\end{eqnarray}
\item $\Delta = 2z$:
\begin{equation}\label{s2z}
src^{(\Delta)} = R^{(2z)} = - 4 \partial_0 \Omega^{I0}_{\ \ I} + 4 \Omega^{I0}_{\ \ I} \Omega^{J0}_{\ \ J} + 2 \Omega_{I0J} \Omega^{I0J} + 2 \Omega_{I0J} \Omega^{J0I}.   
\end{equation}
\item $\Delta = 4-2z$:
\begin{equation}
src^{(\Delta)} = R^{(4-2z)} - \frac{1}{4} F_{AB}^{(2-z)} F^{AB(2-z)} =  \Omega_{IJ0} \Omega^{IJ0} + \Omega_{IJ0} \Omega^{IJ0} \alpha^2. 
\end{equation}
\item $\Delta = \Delta_-$: 
\begin{equation}
src^{(\Delta)} = m^2 \alpha \psi 
\end{equation}
\item $\Delta = \Delta_- + 2$: 
\begin{equation}
src^{(\Delta)} = -\frac{1}{2} F_{AB}^{(1)} F^{AB(1+\Delta_-)} = - (\partial_I \psi + 2 \Omega_{0I}^{\ \ 0} \psi) \Omega_{0}^{\ I0} \alpha. 
\end{equation}
\item $\Delta = \Delta_- + 4-2z$:
\begin{equation}
src^{(\Delta)} = - \frac{1}{2} F_{AB}^{(2-z)} F^{AB(2-z+\Delta_-)} = 2 \Omega_{IJ0} \Omega^{IJ0} \alpha \psi. 
\end{equation}
\item $\Delta = 2\Delta_-$: 
\begin{equation}
src^{(\Delta)} = \frac{1}{2} m^2 \psi^2
\end{equation}
\item $\Delta = 2 \Delta_- +2$: 
\begin{equation}
src^{(\Delta)} = -\frac{1}{4} F_{AB}^{(1+\Delta_-)} F^{AB(1+\Delta_-)} = -\frac{1}{4} (\partial_I \psi + 2 \Omega_{0I}^{\ \ 0} \psi)(\partial_J \psi + 2 \Omega_{0J}^{\ \ 0} \psi) \delta^{IJ}.
\end{equation}
\item $\Delta = 2\Delta_- + 4-2z$:
\begin{equation}
src^{(\Delta)} = - \frac{1}{4} F_{AB}^{(2-z+\Delta_-)} F^{AB(2-z+\Delta_-)} =  \Omega_{IJ0} \Omega^{IJ0} \psi^2. 
\end{equation}
\end{itemize}
We will have contributions from the $src^{(\Delta)}$ term in \eqref{ldelta} only for these values of $\Delta$. 

There is one mixed term in \eqref{qexp} which involves derivatives of the momenta, namely $\nabla_A \pi^A$. This can be written as 
\begin{equation}
\nabla^A \pi_A = \partial^0 \pi_0 - 2 \Omega_{C0}^{\ \ C} \pi^0 + \partial^I \pi_I - 2 \Omega_{CI}^{ \ \ \ C} \pi^I, 
\end{equation}
so a term in $\pi^0$ with dilatation eigenvalue $\Delta$ will contribute a term in $\nabla^A \pi_A$ with eigenvalue $\Delta +z$, while a term in $\pi_I$ with eigenvalue $\Delta$ will contribute a term in $\nabla_A \pi^A$ with eigenvalue $\Delta + 1$.  

We casn now verify that we can solve \eqref{ldelta} for $\lambda^{(\Delta)}$ at each order in terms of the sources and the functional derivatives of $\lambda^{(\Delta')}$ at earlier orders $\Delta' < \Delta$. That is, we want to see that no term coming from $\Delta' \geq \Delta$ can appear in the quadratic term \eqref{qexp} at order $\Delta$. This is not immediately obvious because as noted in \eqref{ltoT} and \eqref{ltopi}, the functional derivatives can give terms with lower $\Delta$, as $\lambda^{(\Delta)}$ determines $T^{0\  (\Delta+1-z)}_{\ I}$ and $\pi_\psi^{(\Delta-\Delta_-)}$. The potential problem with $\pi_\psi$ is easily dealt with; in the explicit quadratic term, the sum only involves $\pi_0^{(s)}$ for $s > \Delta_-$, so the contribution to the quadratic at order $\Delta$ involves $\pi_\psi^{\Delta' - \Delta_-}$ for $\Delta' < \Delta$. $\pi_\psi$ also contributes to $\nabla^A \pi_A$, but this contribution involves $\Delta' + (z-\Delta_-)$, so since $\Delta_- < z$ for all $z$, this involves only $\Delta' < \Delta$. For the $T_{0I}$ piece, when $T_{0I}^{\Delta' + 1-z}$ appears in the quadratic with $K_{0I}^{(s)}$, this will give $\Delta' < \Delta$ if $s < z-1$. For $z < 2$, the smallest eigenvalue in $K_{0I}$ (coming from $\lambda^{(2)}$) is $s = 3-z > z-1$ as $z < 2$. For $z >2$, setting to zero the source terms for $\mathcal E_i$ implies (as we will see below) that the smallest eigenvalue in $K_{0I}$ is $s=z+1$. Thus, all terms in the quadratic piece in \eqref{qexp} are determined by $\lambda^{(\Delta')}$ for $\Delta < \Delta'$, and we can solve \eqref{ldelta} to determine $\lambda^{(\Delta)}$ at each order in terms of the pieces at earlier orders. In particular, this implies that none of the $\lambda^{(\Delta)}$ for $\Delta < z+2$ depend on the undetermined piece $\lambda^{(z+2)}$ which is a non-local function of the boundary data. That is, all the divergences in the naive action \eqref{Dact} are local functions of the boundary values of the bulk fields, determined by solving \eqref{ldelta} recursively. We can then cancel the divergences $\lambda^{(\Delta)}$ for $\Delta < z+2$  by subtracting these local functions of the boundary values of the bulk fields as a local counterterm in our definition of the action.

The solution for the on-shell action in this dilatation expansion for arbitrary values of the source terms will involve dilatation eigenvalues which are the same as the powers of $r$ appearing in the discussion of the solution of the equations of motion in the previous section, namely 
\begin{equation} \label{deltaval}
\Delta = 2n + 2mz + p (4-2z) + q \Delta_-, \quad  n, m, p, q \in \mathbb{N}.
\end{equation}
We have $p=q=0$ if we set the sources for $\mathcal O_\psi$ and $\mathcal E^i$ to zero. There will be additional dilatation eigenvalues which appear in the expansion of the on-shell action if we take a non-zero value for the response function at $\Delta = z+2$, but the terms which are determined in terms of the boundary data will have the eigenvalues \eqref{deltaval}.  

\subsection{Counterterms for $z<2$}

Let us now demonstrate this approach by constructing the divergent pieces of the action for $z <2$, thus identifying the required counterterms.  One would like to give a comprehensive survey of the divergent contributions; unfortunately however, for some values of $z$,  there will be a large number of values of $p$ and $q$ in \eqref{deltaval} which give $\Delta < z+2$, so a comprehensive discussion would be long and not very illuminating. We will therefore confine ourselves to the terms with $p=0$ and $q=0,1$. 

Consider first the terms with $m=n=0$, which don't involve derivatives of the boundary data. The first contribution, at $\Delta=0$, is determined by the leading falloff of the bulk fields. We have the $T^{A\ (0)}_{\ B}$ given by \eqref{Pi0}, which gives the leading term in the on-shell action to be 
\begin{equation}
\lambda^{(0)} = \frac{z T^{0\ (0)}_{\ 0} + T^{I\ (0)}_{\ I}}{z+2} = -2(z+1). 
\end{equation}
Note that this is consistent with re-obtaining $T^0_{\ 0}$ and $T^I_{\ J}$ as the functional derivatives of $\lambda$ using \eqref{Pi}. These values are also consistent with the constraint \eqref{Ksq}, using the value \eqref{s0} for the sources. At $\Delta=\Delta_-$, applying \eqref{ldelta} gives 
\begin{equation}
\lambda^{(\Delta_-)}=  - z \alpha \psi.
\end{equation}
The functional differential equations \eqref{Pi}, \eqref{p} can be checked to be consistently satisfied, with $T^{A(\Delta_-)}_{\ B}  = -z \alpha \psi \delta^A_B$. These first two terms reproduce the no-derivatives counterterm given in \cite{Ross:2009ar} to the relevant order: there we had
\begin{equation}
S_{ct} = 4 + z \alpha \sqrt{-A_\alpha A^\alpha} = 2 (z+1) + z \alpha \psi + \mathcal{O}(\psi^2). 
\end{equation}
It is only the terms at linear order that mattered in \cite{Ross:2009ar}, as there we only considered divergences in the absence of sources.  However, the previous calculation has to be corrected at higher orders. For example, at $\Delta=2\Delta_-$, after a little more calculation, we find
\begin{equation}
\lambda^{(2\Delta_-)} = -\frac{z}{z+1} (2z-1-\Delta_-) \psi^2. 
\end{equation}
This is consistent with the constraint on $\pi_0^{(\Delta_-)}$ in \eqref{pid}. Note that as the frame fields cannot appear undifferentiated in $\lambda$, possible terms in $T^I_{\ 0}$  and $T^0_{\ I}$ arising from functional derivatives of the no-derivative terms will vanish.

Let us now consider the derivative terms. For $\lambda^{(2)}$ there are no quadratic contributions, and putting in the source contributions in \eqref{ldelta} gives 
\begin{equation}
-z \lambda^{(2)} = R^{(2)} - \frac{1}{4} F_{AB}^{(1)} F^{AB(1)} 
\end{equation}
For $\lambda^{(2z)}$, there is also a contribution from the quadratic term involving $\nabla^A \pi_A$, and we have
\begin{equation}
-(2-z)  \lambda^{(2z)} = R^{(2z)} + \frac{1}{2m^2} (\nabla^A \pi_A)^{(z)2} 
\end{equation}
The superscripts $(2)$, $(2z)$ etc indicate that we keep only the term of the stated dilatation eigenvalue in the expression, given explicitly in (\ref{s2},\ref{s2z}). The interesting point to note here is that these expressions are all by construction covariant with respect to coordinate transformations, but because of the different coefficients on the left-hand side, the sum of these two-derivative terms is not covariant with respect to the Lorentz transformations acting on the frame indices. This is as we would expect, given the different scaling dimensions of the frame fields. Put another way, the bulk matter field has singled out a particular direction, so the $0$ and $I$ indices are treated differently. Here we see explicitly how this enters into the required counterterms in the on-shell action. For $z=1$, the two terms will combine into a Lorentz-covariant expression, and this reduces to the usual AdS counterterm.

By taking functional derivatives of these terms, we can evaluate the corresponding components of the stress tensor. For general $\lambda^{(\Delta)}$, the functional variation will be complicated to calculate, but because these low-order terms can be written as the components of a given dilatation eigenvalue in a boundary scalar, we can simply calculate the variation of the scalar quantity and then extract the term of the correct dilatation eigenvalue. For $\lambda^{(2)}$, this gives
\begin{align} 
z T_{00}^{(2)} &= -[2 R_{00} + R + F_{0C} F_0^{\ C} + \frac{1}{4} F^2 - \alpha \nabla_C F^C_{\ 0} ]^{(2)}, \\
z T_{IJ}^{(2)} &= [ -2 R_{IJ} + \delta_{IJ} R - F_{IC} F_I^{\ C} + \frac{1}{4} \delta_{IJ} F^2 ]^{(2)},\nonumber \\
z T_{0I}^{(3-z)} &= -2R_{0I}^{(3-z)}, \nonumber \\
z T_{I0}^{(1+z)} &= [ -2 R_{0I} + \alpha \nabla_C F^C_{\ I}]^{(1+z)}. \nonumber 
\end{align}
While for $\lambda^{(2z)}$, we have
\begin{align}
(2-z)  T_{00}^{(2z)} &= -[ 2 R_{00} + R + \frac{z^2 \alpha}{m^2} \nabla_0 (\nabla^C A_C) ]^{(2z)}, \\
(2-z)  T_{IJ}^{(2z)} &= [ -2 R_{IJ} + \delta_{IJ} R ]^{(2z)}, \nonumber \\
(2-z) T_{0I}^{(z+1)} &= -2R_{0I}^{(z+1)}, \nonumber \\
(2-z) T_{I0}^{(3z-1)} &= [-2R_{0I} - \frac{z^2 \alpha}{m^2} \nabla_I (\nabla^C A_C) ]^{(3z-1)}. \nonumber 
\end{align}

Derivative terms with $q \neq 0$ are more involved; to illustrate this, we consider the term at $\Delta = \Delta_- + 2$, which will also be useful for determining $\pi_\psi^{(2)}$, which will contribute to quadratic terms in the next subsection. This is the first term for which a quadratic term in the stress tensor appears. We have
\begin{equation}
(z - \Delta_-) \lambda^{(\Delta_- + 2)} = - K_{AB}^{(\Delta_-)} T^{AB(2)} + \frac{1}{2} F_{AB}^{(1)} F^{AB(1+\Delta_-)}.
\end{equation}
Using the values of $T_{AB}^{(\Delta_-)}$ and $\pi_0^{(\Delta_-)}$ obtained above to evaluate $K_{AB}^{(\Delta_-)}$, we can rewrite this as 
\begin{equation}
(z - \Delta_-) \lambda^{(\Delta_- + 2)} = \frac{\alpha \psi}{2(z+1)} \left[ z (3z-\Delta_-) T^{00(2)} + z(2z-1-\Delta_-) T^{I \ (2)}_{\ I} + (z+1) \Omega_{IJ0} \Omega^{IJ0} \right]. 
\end{equation}
This gives 
\begin{equation}
(z - \Delta_-) \pi_\psi^{(2)} = \frac{\alpha}{2(z+1)} \left[ z (3z-\Delta_-) T^{00(2)} + z(2z-1-\Delta_-) T^{I \ (2)}_{\ I} + (z+1) \Omega_{IJ0} \Omega^{IJ0} \right]. 
\end{equation}
We see that already at this two derivative level, the form of the counterterms is reasonably complicated. Nonetheless, the conceptual picture is simple: As in the AdS case, the divergent terms are written explicitly in terms of bulk fields evaluated on the boundary. Thus, adding these same expressions to the naive action \eqref{Dact} will cancel the divergences. 

\subsection{Counterterms for $z \geq 2$}

In the asymptotic analysis, we saw that our spacetime would be asymptotically locally Lifshitz for $z \geq 2$ only if we set to zero the sources for the matter operator $\mathcal O_\psi$ and the energy flux $\mathcal E^i$. We will therefore consider here only the counterterms for divergences which arise with those sources set to zero. To consider correlation functions involving $\mathcal O_\psi$ or $\mathcal E^i$, we would need to consider the corresponding sources at the perturbative level, following the interesting recent analysis in the AdS case in \cite{vanRees:2011fr,vanRees:2011ir}. We leave such consideration for future work. 

Setting these sources to zero implies that the dilatation expansion of the action involves only terms with $\Delta = 2n + 2mz$. We want to consider the divergences in the expectation values, and not just the on-shell action, so we should note that setting the source for $\mathcal E^i$ to zero implies that the terms in $T_{0I}$  with $\Delta = 2n+1-z$ will necessarily vanish, as there is no local function of the sources with this dilatation dimension when $\Omega_{IJ}^{\ \ \ 0} = 0$. Thus, in the expansion of $T_{00}, T_{IJ}$ there will only be terms with  $\Delta = 2n + 2mz$, while in the expansion of  $T_{0I}, T_{I0}$  there will only be terms with  $\Delta = 2n +1 + (2m+1)z$. 

Setting the sources for $\mathcal O_\psi$ and $\mathcal E^i$ to zero simplifies the expansion, and we can now discuss all the divergent terms: for the action, the divergent terms have $\Delta = 2n$. We obtained $\lambda^{(0)}$ and  $\lambda^{(2)}$ in the discussion above; we can now consider for example  $\lambda^{(4)}$, which is the only additional term required for $z < 4$. There is no source term for $\Delta = 4$, so the relevant term in the on-shell action is just given by quadratic terms, 
\begin{align}
(z-2) \lambda^{(4)} =& - K_{AB}^{(2)} \pi^{AB (2)} - \frac{1}{2} \pi_A^{(2)} \pi^{A (2)}.
\end{align}
Using the expressions for $T_{AB}^{(2)}$ and $\pi_0^{(2)}$ in the previous subsection, one could compute this counterterm explicitly, but the full expression is unenlightening. What's important about this calculation is not to obtain the explicit forms of the counterterms, but to see that the divergences can be cancelled by adding appropriate counterterms which are local functions of the boundary values of the bulk fields.

\section{Discussion}

In this paper, a geometrically natural definition of asymptotically locally Lifshitz spacetimes was given. This gives boundary data which act as sources for the stress tensor complex of the dual field theory, and can be interpreted as giving a natural non-relativistic curved geometry on the boundary. We have shown that solutions of the equations of motion for arbitrary values of the boundary data exist for $z < 2$ in the context of the massive vector theory. For $2 \leq z < 4$,  bulk solutions satisfying these boundary conditions exist if we set some of the sources to zero. This is natural from the field theory point of view, as the sources we are setting to zero correspond to irrelevant operators. For $z > 4$, however, we would have to set one of the response functions to zero to satisfy these boundary conditions, and we argued in section \ref{lin} that these boundary conditions will have to be appropriately modified. 

For arbitrary boundary data, there are divergences in the bulk action. In section \ref{ct}, we showed that for asymptotically locally Lifshitz boundary conditions, all the divergences are local functions of the boundary values of bulk fields. Thus the divergences can be cancelled by adding appropriate local counterterms to the action. We gave a procedure for explicitly deriving these counterterms, and calculated the first few terms explicitly. The boundary conditions and the construction of counterterms here have a very similar spirit to the familiar AdS calculations; the main message of this work is that dealing with Lifshitz only requires a modest extension of the holographic toolbox.

We should revisit an issue to do with divergences from the previous linearized analysis. In \cite{Ross:2009ar}, there were divergences in $\mathcal E^i$ coming from the other response functions for $z > 2$, and derivative counterterms were introduced to cancel these divergences. In the general analysis we have undertaken here, however, we found that the only divergences in the action came from the terms $\lambda^{(\Delta)}$ for $\Delta < z+2$, and the required counterterm action was uniquely determined as a function of the boundary values of the bulk fields by cancelling these divergences. This is not a contradiction; as observed in \cite{vanRees:2011fr}, the counterterm action is a function of the boundary values of the bulk fields, not of the boundary data. Evaluated on a cutoff surface of finite $r$ and written in terms of the linearised mode solutions, this will have a subleading part involving the response functions. When we consider irrelevant operators, \cite{vanRees:2011fr} showed that some of these subleading terms can be divergent. That is, when we write the divergent part of the action in terms of the linearised modes, it can appear to have a divergence involving the response functions, but when we rewrite it in terms of the boundary values of bulk fields, it depends locally on these fields. These were called pseudo-non-local divergences in \cite{vanRees:2011fr}. Note that the reason that the calculation is simpler here than in \cite{vanRees:2011fr} is that we have switched off the sources for irrelevant operators, so that we can still apply the dilatation expansion technique.

In the approach taken here, the counterterms are necessarily local functions of the boundary values of bulk fields, so they should automatically cancel the divergences seen in \cite{Ross:2009ar}. This can be verified at the two-derivative level by checking that the counterterm needed to cancel $\lambda^{(2)}$ is essentially the same as the two derivative counterterm introduced in appendix A of \cite{Ross:2009ar}\footnote{After correcting a sign error there.}, for the parameter $\sigma_1 = -\frac{1}{z}$. The calculation there then shows that this counterterm will cancel the $k^4$ divergence considered there. 

The analysis given in this paper is essentially complete for $z < 2$. For $z \geq 2$, our restriction to asymptotically locally Lifshitz spacetimes forces us to set sources for irrelevant operators to zero. It may be interesting to analyse the sources for these irrelevant operators perturbatively, following \cite{vanRees:2011fr}; this would be important particularly if there were problems where we wanted to consider finite density for the energy flux $\mathcal E^i$. There is also an issue with the calculation of the expectation value of the scalar operator for $z \geq 2$; the operator becomes irrelevant in this range, but more significantly, the source for this operator appears linearly in the frame fields as well as in the vector field. It would be interesting to see if similar issues arise for the matter content in models which can be embedded in string theory. If so, it would be valuable to understand the correct prescription for calculating this expectation value.

For $z \geq 4$, there is a more substantial issue, as asymptotically locally Lifshitz boundary conditions require that one of the response functions vanishes, leaving us without sufficient freedom to satisfy boundary conditions in the interior of the spacetime. In this regime, our asymptotically locally Lifshitz boundary condition will need to be replaced by a different condition. In section \ref{lin}, we saw that at the linearised level we could either adopt a mixed boundary condition for the frame component $e^{(I)}_t$, which would fix the source for $\mathcal P_i$ as before, or we could continue to impose a simple Dirichlet boundary condition for $e^{(I)}_t$, which would now correspond to fixing $\mathcal P_i$, and regarding it as a source for a vector operator of dimension $z-1$. Understanding the appropriate boundary condition at the non-linear level and extending the analysis performed here to such boundary conditions is perhaps the most important direction for future work here.

\section*{Acknowledgements}

I am grateful for useful discussions with Alex Maloney,  Don Marolf, Mukund Rangamani, and Omid Saremi. This work was supported by STFC; I also thank the Centro de Ciencias de Benasque and the Galileo Galilei Institute for hospitality, and the INFN for partial support during this work. 

\bibliographystyle{utphys}
\bibliography{lifshitz}

\end{document}